\documentclass{sig-alternate}

\usepackage{amsfonts} 
\usepackage{algorithm}
\usepackage{algpseudocode}
\usepackage{epsfig}
\usepackage{subcaption}

\toappear{}

\begin{document}


\algnewcommand{\LineComment}[1]{\State \(\triangleright\) #1}
\algdef{SE}{PForeach}{EndPForeach}[1]{\textbf{pforeach} \(\mbox{#1}\) \textbf{do}}{\textbf{end pforeach}}%

\newcommand{\COMMENT}[1]{}

\newtheorem{defn}{Definition}[section] 
\newtheorem{assum}{Assumption}[section] 
\newtheorem{thm}{Theorem}
\newtheorem{lemma}{Lemma}
\newtheorem{prop}{Proposition}
\newtheorem{cor}{Corollary}

\newcommand*\BlankPage{\newpage\null\thispagestyle{empty}\newpage}


\title{NearBucket-LSH: Efficient Similarity Search\\ in P2P Networks}
\author{Naama Kraus \footnotemark[1] \hspace{0.3in} 
        David Carmel \footnotemark[2] \hspace{0.3in}
				Idit Keidar \footnotemark[1] \footnotemark[2]\hspace{0.3in}
				Meni Orenbach \footnotemark[1]\\ \\
				\large{
				\footnotemark[1] EE Technion, Haifa, Israel \hspace{0.55in} 
				\footnotemark[2] Yahoo Labs, Haifa, Israel
				}
}

\maketitle

\begin{abstract}
We present NearBucket-LSH, an effective algorithm for similarity search in large-scale distributed online social networks organized as peer-to-peer overlays. As communication is a dominant consideration in distributed systems, we focus on minimizing the network cost while guaranteeing good search quality. Our algorithm is based on Locality Sensitive Hashing (LSH), which limits the search to collections of objects, called buckets, that have a high probability to be similar to the query. 
More specifically, NearBucket-LSH employs an LSH extension that searches in near buckets, and improves search quality but also significantly increases the network cost. 
We decrease the network cost by considering the internals of both LSH and the P2P overlay, and harnessing their properties to our needs. 
We show that our NearBucket-LSH increases search quality for a given network cost compared to previous art.
In many cases, the search quality increases by more than $50\%$.
\end{abstract}

\section{Introduction}
\label{sec:intro}

\emph{Online Social Networks (OSNs)} have become popular interaction platforms that serve hundreds of millions of users.
In order to meet scale requirements, commercial OSNs are implemented over a distributed cloud infrastructure.
An alternative paradigm is a \emph{Peer-to-Peer (P2P)} OSN (e.g., ~\cite{peerson09, CutilloMO11, NarendulaPA12, scope2010}),
which offers increased scalability, as well as user privacy, and avoids control by a single authority.

OSN users expose profiles
that reflect their sets of \emph{interests}.
The interest profile may be provided explicitly by the user,
or mined implicitly from her content and activity~\cite{MisloveProfiles,ZhelevaG09,Sharma2012,Adamic01friendsand}.
User \emph{similarity search} is the task of effectively finding OSN users similar to a user query
based on common interests.
It is used for many applications including recommending new friends~\cite{Homophily2001,Xiang2010}, 
as well as for recommending content
based on preferences of similar users~\cite{recommenderSurvey}.
We formally define the similarity search problem in P2P OSNs in Section \ref{sec:model}.
We use the \emph{cosine similarity} function~\cite{Charikar02},
which is a good match for user similarity search in OSNs~\cite{Anderson2012}.
 
A similarity search algorithm in P2P OSNs faces several challenges:
The algorithm should be decentralized in order to fit the P2P architecture. 
As network cost is a dominant consideration in P2P networks,
the algorithm should be network-efficient, while preserving a good search quality.
Furthermore, the similarity search should cope with the dynamic nature of OSNs: 
users join or leave, and users dynamically modify their interest profile.
In this research, we introduce a similarity search algorithm in P2P OSNs that meets these requirements.

We base our algorithm on \emph{Locality Sensitive Hashing (LSH)}~\cite{Gionis99} (see Section \ref{sec:back}),
which is a widespread randomized method
for efficient similarity search in high-dimensional spaces.
LSH hashes an OSN user (based on her interest profile) into a succinct representation,
where the hash values of similar users collide \emph{with high probability (w.h.p.)}.
At a pre-processing stage, LSH maps users into collections of objects called \emph{buckets} based on common hashes.
Upon receiving a query, LSH limits the search to buckets to which the query is mapped;
these contain similar users w.h.p.
LSH improves search time complexity at the cost of search quality,
as the search is approximate and may miss similar users.
In this research, we follow a variant of LSH, called MultiProb-LSH~\cite{Lv07},
which increases search quality by additionally searching \emph{near buckets},
i.e., buckets similar to the query's bucket.

In Section \ref{sec:alg} we present our \emph{NearBucket-LSH} algorithm, 
which integrates LSH into a P2P architecture.
For our P2P overlay we use \emph{Content Addressable Network (CAN)}~\cite{CAN01},
which is a good fit for a distributed LSH implementation, as we later show.
We use CAN to dynamically map and store LSH buckets within nodes,
and refresh bucket contents once in a while in order to adjust to changes in the data. 
Upon search, we use CAN to locate the buckets to search in.

In P2P settings, searching additional buckets entails contacting additional nodes,
which is a network-costly operation.
We improve the network-efficiency when searching near buckets by exploiting the internals of CAN:
We observe that in CAN, near buckets reside in a bucket's neighboring nodes,
and thus contacting them incurs a low network cost.
We further eliminate this network cost by caching near buckets in each CAN node.

In Section \ref{sec:analysis}, we analytically study NearBucket-LSH for the cosine similarity metric. 
We first prove that for any fixed number, $k$, NearBucket-LSH's choice of $k$ near buckets to search in is optimal. 
We next compare NearBucket-LSH to LSH, as well as to Layered-LSH~\cite{Haghan09,Bahmani12}, 
a previously suggested LSH variant for distributed systems,  
which also searches near buckets with the goal of reducing network cost. 
Our analysis shows that NearBucket-LSH achieves better success probability for a given network cost than the other two approaches.

In Section \ref{sec:eval}, we provide an empirical evaluation of our algorithm
using three real world OSN datasets: DBLP, LiveJournal, and Friendster~\cite{Yang2012};
Friendster is the largest of the three, having $7,\!944,\!949$ users, and $1,\!620,\!991$ interests.
We first empirically reproduce the theoretical results of Section \ref{sec:analysis}.
We then measure search quality according to two metrics: recall and precision.
Our experiments demonstrate that for all three datasets and all metrics,
the cache-based NearBucket-LSH provides the greatest search quality for a given network cost,
compared to LSH and Layered-LSH.
For example, in LiveJournal, 
for an average network cost of $96$ messages per query, 
NearBucket-LSH increases recall by more than $50\%$
compared to LSH and Layered-LSH, and improves precision from $0.59$ to $0.87$. 	

In summary, the contributions of this paper are:
\begin{itemize}
\item a P2P OSN \textbf{similarity search algorithm}, NearBucket-LSH,
which offers a strong connection between the search quality aspects of LSH
and the distributed aspects of the underlying infrastructure (CAN in our case);
\item a \textbf{formal analysis} of NearBucket-LSH for the cosine similarity metric, 
showing its optimal choice of near buckets and superior success probability for a given network cost compared to previous art; and
\item an extensive \textbf{empirical evaluation} using three large real-world OSN datasets, 
confirming our analysis and extending it to standard search quality metrics.
\end{itemize}
Section \ref{sec:conc} concludes the paper and suggests some directions for future research.

\section{Model and Problem Definition}
\label{sec:model}
In this section we detail the model we consider.
We formally define the notion of similarity search (Section \ref{sec:ssdef}),
and provide details about P2P OSNs (Section \ref{sec:p2p}).  

\subsection{User Similarity Search in OSN}
\label{sec:ssdef}
Each OSN user exposes an \emph{interest profile} (either explicitly or implicitly), 
which we represent as a non-negative weighted feature vector in a high $d$-dimensional vector space $V=(\mathbb{R}^+_0)^d$
(in the largest dataset we experiment with $d=1,\!620,\!991$).
We use $v_i$ to denote the $i$-th entry of vector $v$
(corresponding to the $i$-th interest feature).
The interests-weighting scheme may be arbitrary.
A \emph{similarity function}~\cite{Chierichetti12} measures the similarity between two user vectors:
\begin{defn}[similarity function]
A similarity function is a symmetric function $S:V^2 \rightarrow [0,1]$
such that $\forall u,v \in V,
S(u,v) = S(v,u)\ and \ S(v,v)=1$.
\end{defn}
The similarity function returns a \emph{similarity value} within the range $[0,1]$,
where a similarity value of $1$ denotes complete similarity,
and $0$ denotes no similarity.

An \emph{$m$-similarity search} algorithm accepts as an input a \emph{query} vector $q \in V$.
It returns a unique \emph{ideal result set} of $m$ user vectors that are most similar to $q$,
according to the given similarity function.
Computing the ideal result set is not always desired, as it may be inefficient.
An \emph{approximate $m$-similarity search} algorithm trades-off efficiency with accuracy.
Given a query $q$, it returns an \emph{approximate result set} of $m$ user vectors,
which may differ from $q$'s ideal result set. 

A commonly used similarity function is the \emph{cosine similarity},
also proposed in the context of similarity between OSN users~\cite{Anderson2012}.
The similarity between two vectors $u,v \in V$ is defined as the cosine of the angle between them:
\begin{equation}
\label{eq:cos}
sim_{\cos}(u,v) = \frac{u \cdot v}{\left\|u\right\| \cdot \left\|v\right\|}.
\end{equation}
Yet, LSH was not defined for cosine similarity,
but rather, for the closely related \emph{angular similarity}~\cite{Charikar02}:
\begin{equation}
\label{eq:ang}
sim_{ang}(u,v) = 1-\frac{\theta(u,v)}{\pi},
\end{equation}
where $\theta(u,v)$ is the angle between $u$ and $v$.
As the angular and cosine similarities are closely related,
we can similarly analyze LSH for cosine similarity~\cite{Charikar02,Chierichetti12}.

\subsection{P2P OSN and CAN}
\label{sec:p2p}
P2P networks are distributed systems organized as overlay networks with no central management.
Nodes (also called \emph{peers}) are autonomous entities that may join or leave at any time;
content is distributed among the participating nodes.
P2P networks provide massive scalability, fault tolerance, privacy, anonymity, and load balancing
(see~\cite{P2PSurvey05} for a survey).
We consider a P2P Online Social Network~\cite{peerson09, CutilloMO11, NarendulaPA12, scope2010},
in which users' content is distributed among nodes.
Any node in the P2P OSN may initiate a similarity search query.
Typical OSNs include hundreds of millions of users, and millions of interest features.
We consider a dynamic data model, in which users join or leave the OSN
and existing users update their interest profiles.
We assume the update rate is several orders of magnitude lower than the query rate
($5-10$ orders of magnitude, depending on the specific application).

In our algorithm, we use CAN~\cite{CAN01} as our overlay,
which naturally fits a distributed LSH implementation, as we later show.
CAN implements a self-organizing P2P network
representing a virtual $c$-dimensional Cartesian coordinate space on a $c$-torus.
The Cartesian space is dynamically partitioned into \emph{zones},
which are distributed among CAN nodes.
CAN implements a \emph{Distributed Hash Table (DHT)} abstraction, 
which provides a distributed \emph{lookup} operation that accepts a vector as key,
and returns a node that owns the zone to which the vector belongs.
Each node maintains a table of \emph{neighbors}, which are nodes that own zones adjacent to its own.
These tables are used for routing messages within CAN.

\section{Background and Previous Work}
\label{sec:back}
Before diving into our algorithm, we provide essential background and overview previous work.
We start in Section \ref{sec:back_lsh}, by overviewing LSH.
In Section \ref{sec:back_mpLSH} we present MultiProb-LSH,
a centralized LSH extension that improves search quality.
Section \ref{sec:back_layLsh} discusses Layered-LSH~\cite{Haghan09,Bahmani12}, 
which is a distributed LSH implementation that optimizes network cost.
In addition to distributed solutions as discussed in Section \ref{sec:back_layLsh}, 
there are also parallel LSH variants, e.g.~\cite{PLSH}.
However, these do not focus on improving network-efficiency, which is not of essence in a parallel setting.

\subsection{Locality Sensitive Hashing}
\label{sec:back_lsh}
Locality sensitive hashing (LSH) is a widely used probabilistic method 
that tackles the efficiency challenge of similarity search in high dimensional spaces.
Given a similarity measure, LSH maps a vector in a high dimensional space
into a representation in a lower dimensional space,
so that the probability of two vectors to collide equals their similarity under the given measure.
 
Charikar~\cite{Charikar02} formally defines LSH as follows:
\begin{defn}[LSH]
A locality sensitive hashing\\ scheme is a distribution on a family $\mathcal{H}$ of hash functions operating on a
collection of vectors, so that for two vectors u, v,
\begin{equation}
Pr_{h \in \mathcal{H}}\left[h(u)=h(v)\right]=sim(u,v),
\end{equation}
where sim(u,v) is some similarity defined on the collection of vectors.
\end{defn}

Charikar~\cite{Charikar02} proposes an LSH hash family $LSH_{ang}:V \rightarrow \left\{0,1\right\}$ for \emph{angular similarity},
which is based on Goemans and Williamson's~\cite{Goemans95} random hyper-plane rounding technique.
Charikar's LSH may be further used under the cosine similarity function~\cite{Charikar02, Chierichetti12}.

Gionis et al.~\cite{Gionis99} introduce an approximate similarity search algorithm based on LSH.
In a pre-processing stage, they partition the data vectors into buckets according to their LSH hash values.
Given a query vector, the algorithm computes its hash and searches vectors in the corresponding bucket. 
The algorithm is resource-efficient as it searches over a subset of the data. 
It provides good search quality,
as it selects vectors that have a high probability to be similar to the query.
However, it may suffer from low recall, 
as similar items to the given query may be mapped (though not with high probability) to non-searched buckets.

The LSH algorithm is parametrized by $k$ and $L$, where $k << d$, the vector dimension.
In order to increase precision, 
the algorithm defines a family $\mathcal{G}$ of hash functions,
where each $g(v) \in \mathcal{G}$ is a concatenation of $k$ functions chosen randomly and independently from $\mathcal{H}$.
In the case of angular similarity, $g: V \rightarrow \left\{0,1\right\}^k$,
i.e., $g$ hashes $v$ into a binary \emph{sketch vector}, which encodes $v$ in a lower dimension $k$.

For two vectors $u,v$, $Pr_{g \in \mathcal{G}}\left[g(u)=g(v)\right]=(sim(u,v))^k$,
for any randomly selected $g \in \mathcal{G}$.
The larger $k$ is, the higher the precision. 
In order to increase the recall, 
the $LSH$ algorithm selects $L$ functions randomly and independently from $\mathcal{G}$.
The data is now replicated in $L$ hash tables, where each vector is mapped to $L$ buckets. 
Upon query, search is performed in $L$ buckets, which increases recall at the cost of additional storage and processing.

\subsection{MultiProb-LSH}
\label{sec:back_mpLSH}
Increasing $L$ in order to improve recall increases the storage size,
due to increasing the number of hash tables.
In order to improve search quality without increasing the storage cost, 
Entropy-LSH~\cite{Panigrahy06} and MultiProb-LSH~\cite{Lv07},
search in additional buckets within the same hash table.
MultiProb-LSH is motivated by the observation that buckets that slightly differ from the query's \emph{exact bucket} $g(q)$,
have a high probability to contain vectors similar to the query.
Searching in such near buckets yields additional similar results w.h.p., 
which increases recall for a given $L$.
MultiProb-LSH was introduced in the context of the $l_p$ norm;
here, we apply its principles to cosine-based LSH, in the context of P2P OSN.

\subsection{Layered LSH}
\label{sec:back_layLsh}
In P2P networks, contacting a near bucket involves performing a DHT lookup of the bucket's node, 
which incurs high network cost.
Haghani et al.~\cite{Haghan09} and Bahmani et al.~\cite{Bahmani12} tackle this network-efficiency problem using 
the concept of \emph{Layered-LSH}.
Layered-LSH maps buckets to nodes using a second LSH,
which is defined over the buckets' sketch vectors and
captures similarity between buckets.
Queries now access a single node holding a bucket of buckets,
which reduces the network cost significantly.
Layered-LSH was originally presented in the context of the $l_p$ norm.
In Section \ref{sec:laylsh}, we show that in the case of cosine similarity,
Layered-LSH is equivalent to the basic LSH for an appropriate choice of $k$.

Layered-LSH sometimes searches in additional nodes, 
essentially implementing a distributed variant of MultiProb-LSH\footnote{Bahmani et al.~\cite{Bahmani12} 
present their Layered-LSH algorithm in the context of Entropy-LSH.
Nevertheless, as the authors indicate, their Layered-LSH algorithm is also applicable when combined with MultiProb-LSH.},
which improves search quality, though, again, incurs an extra network cost for the additional DHT lookups.
Haghani et al.~\cite{Haghan09} propose a heuristic that linearly searches close nodes
along the DHT links.
This method is not applicable to the cosine-similarity case, 
due to the differences in the sketch vector representations.

\section{Algorithm}
\label{sec:alg}
Our algorithm is based on locality sensitive hashing,
reviewed in Section \ref{sec:back} above.
In order to implement our P2P user similarity search,
we construct a dedicated overlay above the CAN infrastructure.
We distribute LSH buckets of user vectors among the overlay nodes,
occasionally refresh their content to adjust for changes,
and route search queries to the appropriate buckets, as described in Section \ref{sec:canLSH}.
In Section \ref{sec:nblsh}, we extend this basic approach to also search in near buckets.

\subsection{CAN-based LSH}
\label{sec:canLSH}
\paragraph*{The Overlay}
We use a $k$-dimensional CAN (i.e., $c=k$) to store and lookup LSH buckets in a decentralized manner.
For simplicity, we assume that $N=2^k$, where $N$ denotes the number of CAN nodes.
Note that our overlay may be formed by a subset of the OSN nodes,
but for simplicity of the description, we assume all OSN nodes participate in the overlay.
Each CAN node owns the zone of a single $k$-dimensional binary vector $v$ representing some LSH sketch vector,
and maintains the bucket of user vectors that are mapped to $v$ by some hash function $g \in \mathcal{G}$.
We name such a node the \emph{bucket node} of $v$.
The bucket node provides a local similarity search facility over its locally stored user vectors.
The local search time is typically proportional to the searched bucket size~\cite{Gionis99}.
The internal bucket data-structure and local search implementation are orthogonal to this research.

Each CAN node in our overlay has $k$ neighbors;
the $i$-th neighbor of node $v$ owns a vector $u$ that differs from $v$ in the $i$-th entry only.
Routing a message from node $v$ to one of its neighbors requires a single hop,
i.e., a single message. 
Routing a message from an arbitrary source node $v$ to an arbitrary target node $u$,
entails modifying the binary vector entries that differ between $u$ and $v$.
Two vectors of length $k$, differ in $k/2$ entries,
and thus, the expected path length is $k/2$ hops\footnote{
Note that in a general $c$-dimensional CAN of $N$ nodes, the expected routing length is $c/4\left(N^{1/c}\right)$~\cite{CAN01},
which equals $k/2$ for $c=k$ and $N=2^k$.}.

The $L$ hash functions $g=\left\{g_1, \cdots, g_L\right\}$ 
are randomly selected from $\mathcal{G}$ a priori.
They are given to the distributed algorithm as a configuration parameter,
and are known to all bucket nodes.
CAN supports multiple hash functions~\cite{CAN01}, 
which we use for supporting multiple $g_i$'s
and mapping each user vector into $L$ bucket nodes.

\paragraph*{Bucket Maintenance}
Our algorithm constructs and refreshes the buckets continuously, in a decentralized manner.
Thus, each bucket node stores soft state that is regularly refreshed.
Each user periodically re-hashes its vector using LSH
into $L$ sketch vectors in $\left\{1,0\right\}^k$.
It then performs DHT lookups to locate the corresponding bucket nodes, and sends them the fresh user vector.
Note that the user vector may or may not have changed since the previous update message.

We do not construct buckets a priori.
Rather, bucket construction is triggered by vector update messages.
A CAN node becomes an active bucket node when it first receives a notification of some user vector.
Since user vectors change dynamically, their hashes change accordingly.
Obsolete vectors that are not refreshed for a certain predefined length of time are garbage-collected from bucket nodes.

\paragraph*{Query Processing}
Each P2P node may trigger an $m$-similarity search request for an input query $q$.
The similarity search follows the LSH algorithm~\cite{Gionis99}, using our overlay.
The initiating node, denoted $n$, 
activates the function \textsc{Query} in Algorithm \ref{alg:dLSH}:
It hashes $q$ into $L$ sketch vectors according to the pre-defined $g_i$ functions,
looks-up $L$ corresponding bucket nodes $n_i$ using CAN,
and sends $m$-similarity search requests with the input query $q$ to all $L$ bucket nodes in parallel.
Each bucket node locally performs an $m$-similarity search (function \textsc{SimSearch} in Algorithm \ref{alg:dLSH}),
and sends back a set of up to $m$ results, 
associated with their similarity values.
Node $n$ receives $L$ result sets, which it merges and sorts according to the similarity values.
It then returns a final $m$-result set to the caller.

\begin{algorithm}
\caption{Distributed LSH Algorithm}
\label{alg:dLSH}
\begin{algorithmic}[1]
\Function{Query}{$q$} \Comment{At the query node}
\label{func:query}
	\PForeach{$g_i \in g$} \Comment{A parallel foreach}
	\label{line:pforeach}
	  \State $v_i \gets g_i(q)$
		\State $n_i \gets DHT.\Call{lookup}{v_i}$ \Comment{Lookup bucket node}
		\label{line:dhtLookup}
		\State $n_i.\Call{SendReq}{SimSearch, q, n}$ \Comment{Send request}
	\EndPForeach
	 \State $hits \gets \mbox{collect results from bucket nodes}$
	\State \Return {top $m$ hits} \Comment{Rank and return top $m$}
	\label{line:merge1}
\EndFunction
\Statex

\Function{SimSearch}{$q$, $n$} \Comment{Query $q$ from $n$}
\label{func:localSS}
	\State $res \gets Bucket.\Call{LocalSimSearch}{q}$ \Comment{Local search}
	\State $n.\Call{SendRes}{res}$ \Comment{Send back result}
\EndFunction
\Statex

\end{algorithmic}
\end{algorithm}

\subsection{NearBucket-LSH}
\label{sec:nblsh}
Given a query $q$ and some hash function $g \in \mathcal{G}$, 
the basic LSH algorithm searches in the exact bucket $g(q)$.
NearBucket-LSH extends LSH to also search in near buckets that differ from $g(q)$ in exactly one vector entry,
i.e., one bit is flipped.
As we analytically show in Section \ref{sec:analysis}, searching in near buckets increases the probability
to find similar users.

Contacting a neighbor costs a single message, for a total of $kL$ messages per query.
We further eliminate these additional messages by caching $k$ near buckets at each CAN node.
In order to maintain fresh caches, each node periodically sends its bucket to its neighbors.
The cache requires an additional storage of size $kB$ at each node,
where $B$ is the average bucket size.
Note that our cache is only used for storing near buckets.

A CAN node maintains a table of $k$ neighbors that differ from it in exactly one entry, 
which are also the neighbors that hold the desired near buckets.
Given a query $q$,
NearBucket-LSH uses a query function similar to the function \textsc{Query} in Algorithm \ref{alg:dLSH}:
to contact the $L$ exact bucket nodes using CAN.
But here, the sent request is SimSearchNB.
Once such a request reaches some exact bucket node, 
it activates the function \textsc{SimSearchNB} in Algorithm \ref{alg:dNBLSH}:
The node first performs a local similarity search in its own bucket (line \ref{line:localExact} in Algorithm \ref{alg:dNBLSH}).
Then for each of its $k$ neighbor nodes $n_j$,
$j=\left\{1,\cdots,k\right\}$ ,
it checks if that node's bucket is cached locally.
If it is, it searches it,
and if not, it forwards the query to that node.
In case messages are forwarded,
bucket nodes perform local $m$-similarity searches of query $q$ in parallel,
and each returns a result set to the initiating node $n$.

\begin{algorithm}
\caption{Distributed NearBucket-LSH Algorithm}
\label{alg:dNBLSH}
\begin{algorithmic}[1]
\Function{SimSearchNB}{q, n} \Comment{Query $q$ from $n$}
\label{func:ssNB}
	\State $res \gets Bucket.\Call{LocalSimSearch}{q}$ \Comment{Local search}
	\label{line:localExact}
	\State $n.\Call{SendRes}{res}$ \Comment{Send back result}
	\PForeach{$j \in \left\{1,\cdots,k\right\}$} \Comment{A parallel foreach}
		\State $n_j \gets Neighbors.j$ \Comment{Extract the $j$-th neighbor}
		\If{$Bucket_j.isCached$}
		\label{line:cache}
		\State $res \gets Bucket_j.\Call{LocalSimSearch}{q}$ \Comment{Local search}
	  \State $n.\Call{SendRes}{res}$ \Comment{Send back result}
		\Else
		\State $n_j.\Call{SendReq}{SimSearch, q, n}$ \Comment{Forward request}
		\label{line:forward}
		\EndIf
	\EndPForeach	
\EndFunction
\Statex

\end{algorithmic}
\end{algorithm}

It is possible to cache all $k$ near buckets or any subset of them.
For the purpose of the analysis and evaluation in the next sections,
we refer to the following two extremes:
we name NB-LSH a NearBucket-LSH that does not use caching at all,
and CNB-LSH a NearBucket-LSH that caches all $k$ near buckets.

\section{Theoretical Analysis}
\label{sec:analysis}
We now theoretically analyze the algorithm's capability of retrieving similar objects.
In Section \ref{sec:succProb} we formulate the probability of an algorithm to successfully find 
an object with some given similarity to the query.
We do this for both LSH and NearBucket-LSH.
In Section \ref{sec:laylsh} we explain how one could adapt Layered-LSH to use cosine similarity,
and show that in this context, Layered-LSH is equivalent to NearBucket-LSH for some choice of parameters.
In Section \ref{sec:succComp}
we compare the algorithms,
finding that NearBucket-LSH guarantees a greater success probability for a given network cost than the other approaches.

\subsection{Success Probability Formulation}
\label{sec:succProb}
The basic building block in our analysis
is the \emph{success probability}~\cite{Lv07} of an algorithm $A$ to find object $y$ that has a similarity value $s$ 
to query object $x$, under a random selection of $g \in \mathcal{G}$.
We denote this success probability by $SP(A,s)$.
Note that the meaning of success is that the algorithm searches within a bucket containing $y$, 
hence $y$ is found. 
This does not necessarily imply, however, that $y$ will be retrieved,
as the algorithm might find other items that are more similar to $x$.

We use angular-LSH~\cite{Charikar02} in our analysis, 
but first show how to convert it to cosine similarity.
Let $s$ denote the angular similarity between vectors $x$ and $y$.
Let $t$ denote the cosine similarity between the same vectors.
From Equation \ref{eq:ang}:
\begin{equation}
\label{eq:costoang}
s = 1-\frac{arccos(t)}{\pi}.
\end{equation}
Let $\theta$ denote the angle between $x$ and $y$.
As we consider non-negative vectors, it holds that $\theta \in [0,\frac{\pi}{2}]$. 
This implies (by the definitions of angular and cosine similarities in Section \ref{sec:ssdef}),
that the angular similarity $s$ satisfies $s \in [0.5,1]$,
and the cosine similarity $t$ satisfies $t \in [0,1]$.
In particular, for orthogonal vectors, the cosine similarity is $0$ and the angular similarity is $0.5$.

We denote the angular-similarity LSH algorithm with parameters $k$, $L$ as $LSH(k,L)$.
According to the LSH theory~\cite{Charikar02}, for a randomly selected $h \in \mathcal{H}$:
\begin{equation}
\label{prop:h_eq}
Pr_{h \in \mathcal{H}}\left[h(x)=h(y)\right]=s,\mbox{ and }Pr_{h \in \mathcal{H}}\left[h(x) \ne h(y)\right]=(1-s).
\end{equation}

For a randomly selected $g \in \mathcal{G}$, which is a concatenation of $k$ $h_i \in \mathcal{H}$ hash functions,
it follows that for an angular LSH algorithm that searches in one exact bucket, $LSH(k,1)$: 
\begin{equation}
\label{prop:exBucket}
SP(LSH(k,1),s)=s^k.
\end{equation}

If $L$ randomly selected hash functions $g_i \in \mathcal{G}$ are in use, 
then each object is mapped to $L$ buckets.
An angular LSH algorithm that searches in $L$ exact buckets may find $y$ in any of these buckets:
\begin{prop}
\label{prop:lsh}
\[
SP(LSH(k,L),s)=1-\left(1-s^k\right)^L.
\]
\end{prop}

We now move on to analyze search in near buckets.
We first consider a near bucket of (exact) bucket $g(x)$, which differs from $g(x)$ in exactly one entry.
Following Equation \ref{prop:h_eq},
the success probability of a near bucket of $g(x)$ is:
\begin{equation}
\label{prop:nbBucket}
s^{k-1}(1-s).
\end{equation}

As we previously indicated, $s \in [0.5,1]$.
This implies that $\forall s, (1-s) \le s$, and therefore, $s^{k-1}(1-s) \le s^k$.
Following Equations \ref{prop:exBucket} and \ref{prop:nbBucket}:
\begin{prop}
\label{prop:nearVSexact}
The success probability when searching in an exact bucket
is greater or equal to the success probability when searching in a near bucket.
\end{prop}

We next generalize to buckets that differ from $x$'s exact bucket in $0 \le b \le k$ entries.
We name such buckets \emph{$b$-near buckets}
(note that a $0$-near bucket is an exact bucket).
Similarly to Proposition \ref{prop:nearVSexact}, the success probability of a $b$-near bucket of $g(x)$ is:
\begin{equation}
\label{prop:bnbBucket}
s^{k-b}(1-s)^b.
\end{equation}
As $s \in [0.5,1]$, it follows that for $0 \le b_1 < b_2 \le k$, 
$s^{k-b_2}(1-s)^{b_2} \le s^{k-b_1}(1-s)^{b_1}$.
We can now generalize Proposition \ref{prop:nearVSexact}:
\begin{prop}
\label{prop:bNearVS}
The success probability when searching in a $b_1$-near bucket is
greater or equal to the success probability when searching in a $b_2$-near bucket,
for any $0 \le b_1 < b_2 \le k$. 
Hence, NearBucket-LSH's selection of $k$ $1$-near buckets is optimal,
with respect to any other $k$ buckets selected for search, in addition to the exact bucket.
\end{prop} 

The exact bucket and its near buckets are disjoint, 
as an object is mapped to exactly one bucket according to a specific $g$.
Thus, the success probability to find $y$ in an exact bucket 
or in its $k$ near buckets is a union of disjoint events.
NearBucket-LSH searches in $1$-near buckets,
therefore, the success probability of NearBucket-LSH for $L=1$ is:
\[
SP(\mbox{NearBucket-LSH}(k,1), s)=s^k + ks^{k-1}(1-s).
\]

NearBucket-LSH searches in $L$ exact buckets each along with its $k$ $1$-near buckets.
Thus,
\begin{prop}
\label{prop:nblsh}
\[
SP(\mbox{NearBucket-LSH}(k,L),s)=1-(1-(s^k + ks^{k-1}(1-s)))^L.
\]
\end{prop}

\subsection{Layered-LSH}
\label{sec:laylsh}
Layered-LSH was introduced in the context of the $l_p$ norm.
We next explain how to adapt it to cosine similarity and show that in this context,
Layered-LSH is equivalent to the basic LSH.
Recall that Layered-LSH maps near buckets to the same node w.h.p.
According to Proposition \ref{prop:bNearVS}, 
in our case this means mapping buckets that differ in a small number of entries to the same node.
This can be achieved by using Hamming-based LSH~\cite{Gionis99,Chierichetti12}.

Let $g_{cos}$ be the cosine-LSH used for mapping vectors to buckets.
By definition, $g_{cos}$ is a concatenation of $h_i$ cosine-based LSH functions chosen randomly and independently from $\mathcal{H}$.
Let $g_{ham}$ be the Hamming-LSH used for mapping buckets to nodes.
Hamming-based LSH hashes a binary vector to another binary vector of a lower dimension $k$, 
by randomly and independently selecting $k$ entries of the input vector.
In our case, this resorts to randomly and independently selecting $k$ entries from $g_{cos}(v)$,
each of which corresponds to some $h_i \in \mathcal{H}$.
We get that $g_{ham}(g_{cos}(v))$ maps $v$ to a node according to $k$ randomly selected $h \in \mathcal{H}$ functions,
which is equivalent to using the cosine-based LSH with parameter $k$.

\subsection{Success Probability Comparison}
\label{sec:succComp}
We now use Propositions \ref{prop:lsh} and \ref{prop:nblsh} and Equation \ref{eq:costoang}
to compare the success probabilities of LSH, Layered-LSH, and NearBucket-LSH, as a function of different parameters.
As Layered-LSH is equivalent to LSH, we refer to both as LSH in this discussion.

\begin{figure*}[hbt]
        \centering
        \begin{subfigure}{0.3\textwidth}  
                \includegraphics[height=1.3in]{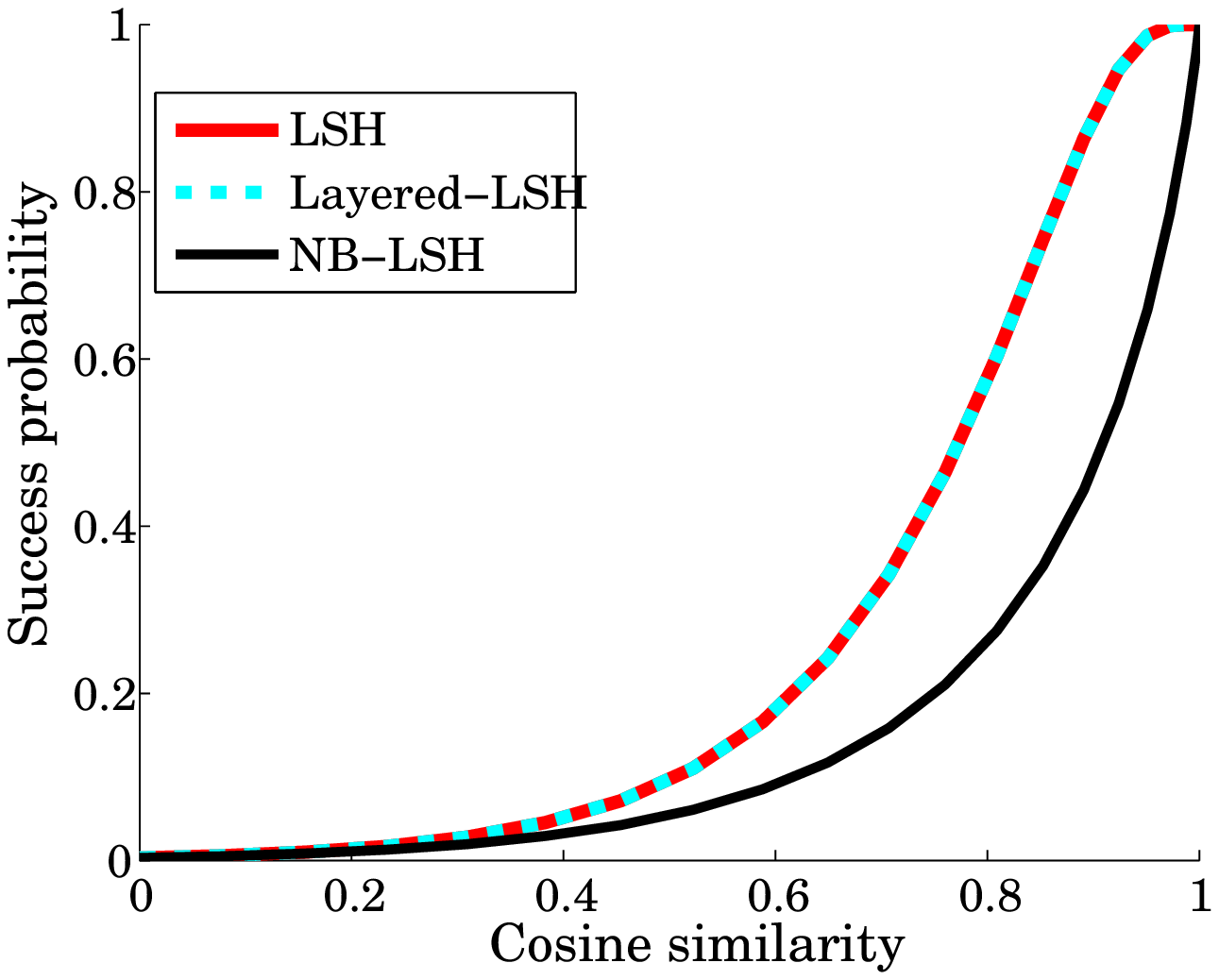} 
                \caption{searched buckets: 13}
        \end{subfigure}%
        ~ 
        \begin{subfigure}{0.3\textwidth}
                \includegraphics[height=1.3in]{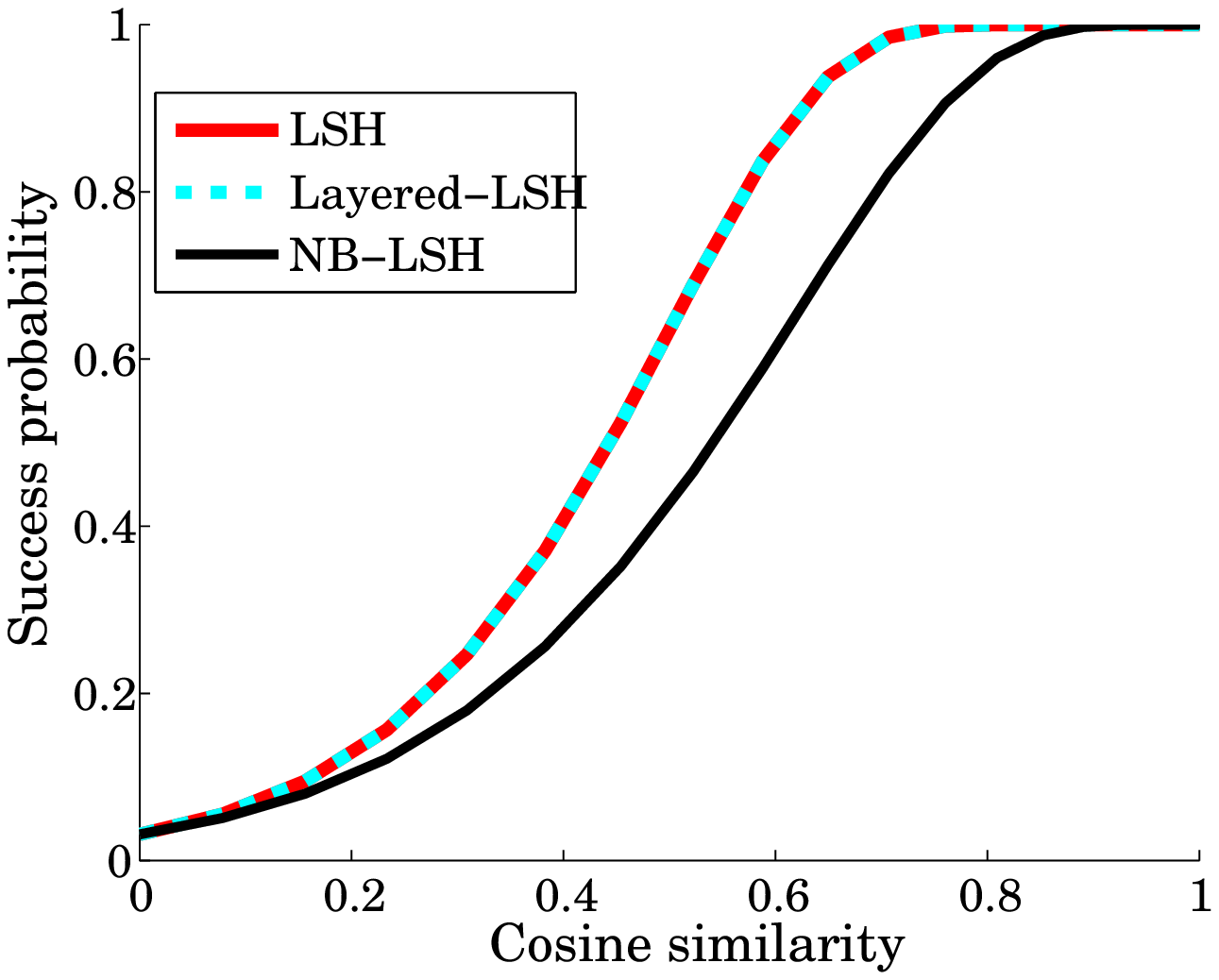}
                \caption{searched buckets: 130}
        \end{subfigure}
				~ 
        \begin{subfigure}{0.3\textwidth}
								\includegraphics[height=1.3in]{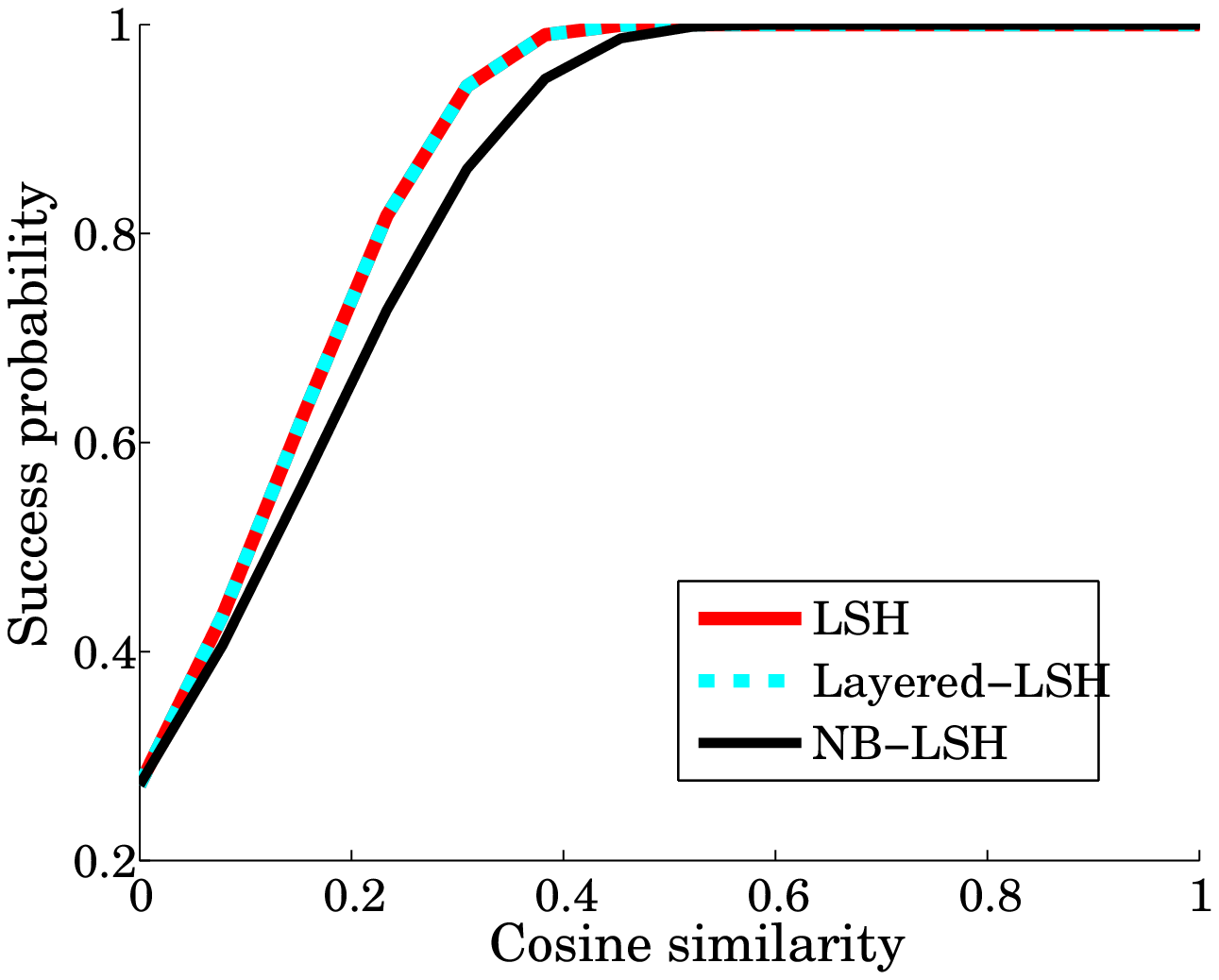}
                \caption{searched buckets: 1300}
        \end{subfigure}
        \caption{Analytical success probability as a function of the number of searched buckets ($k=12$).
				         LSH guarantees a greater or equal success probability compared to NearBucket-LSH,
								 as it searches in exact buckets only. The gap decreases as the number of buckets increases.}
				\label{fig:analysis_numBuckets}
\end{figure*}

\paragraph*{Constant Number of Buckets Searched}
Both LSH and NearBucket-LSH select a subset of the buckets to search in.
Given $k$ and $L$, LSH searches in $L$ exact buckets,
whereas NearBucket-LSH searches in $L$ exact buckets plus $kL$ near buckets.
Figure \ref{fig:analysis_numBuckets} compares the success probabilities of LSH and NearBucket-LSH for a constant number of searched buckets.
For the purpose of this demonstration,
we selected $k=12$ and corresponding $L$ values that entail searching in $13$, $130$, and $1300$ buckets.
Notice that LSH's success probability is greater than or equal to that of NearBucket-LSH
for any constant number of searched buckets.
This stems from the fact that LSH searches in exact buckets only, 
while NearBucket-LSH searches in both exact and near buckets.
As we have shown in Proposition  \ref{prop:nearVSexact}, 
searching in exact buckets yields a greater success probability compared to searching in near buckets.
We later show that when considering network efficiency,
search in exact buckets incurs a high network cost, 
which makes exact bucket search less desirable. 

\begin{figure*}[hbt]
        \centering
        \begin{subfigure}{0.3\textwidth}
                \includegraphics[height=1.3in]{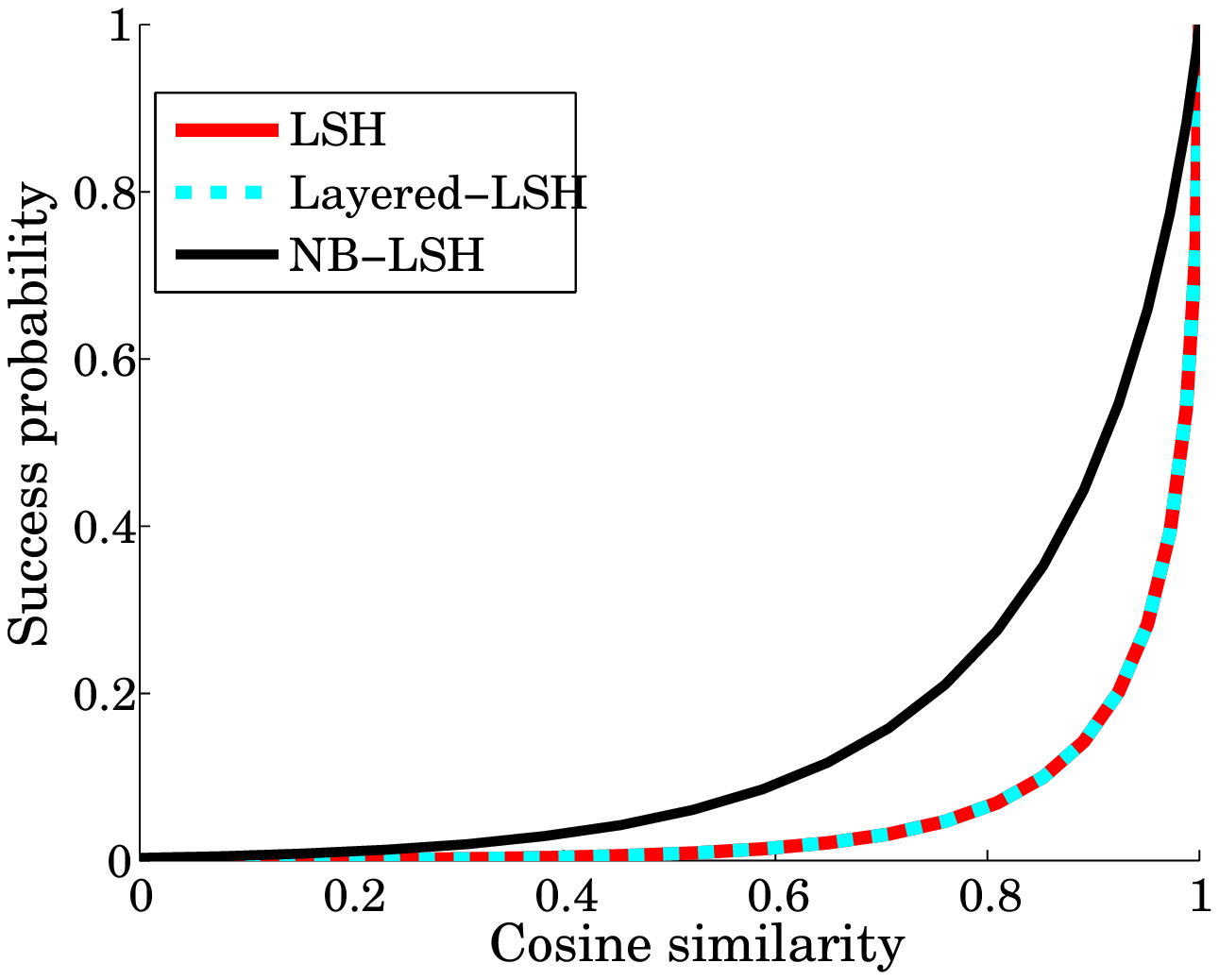}
                \caption{$L=1$}
        \end{subfigure}%
        ~ 
        \begin{subfigure}{0.3\textwidth}
                \includegraphics[height=1.3in]{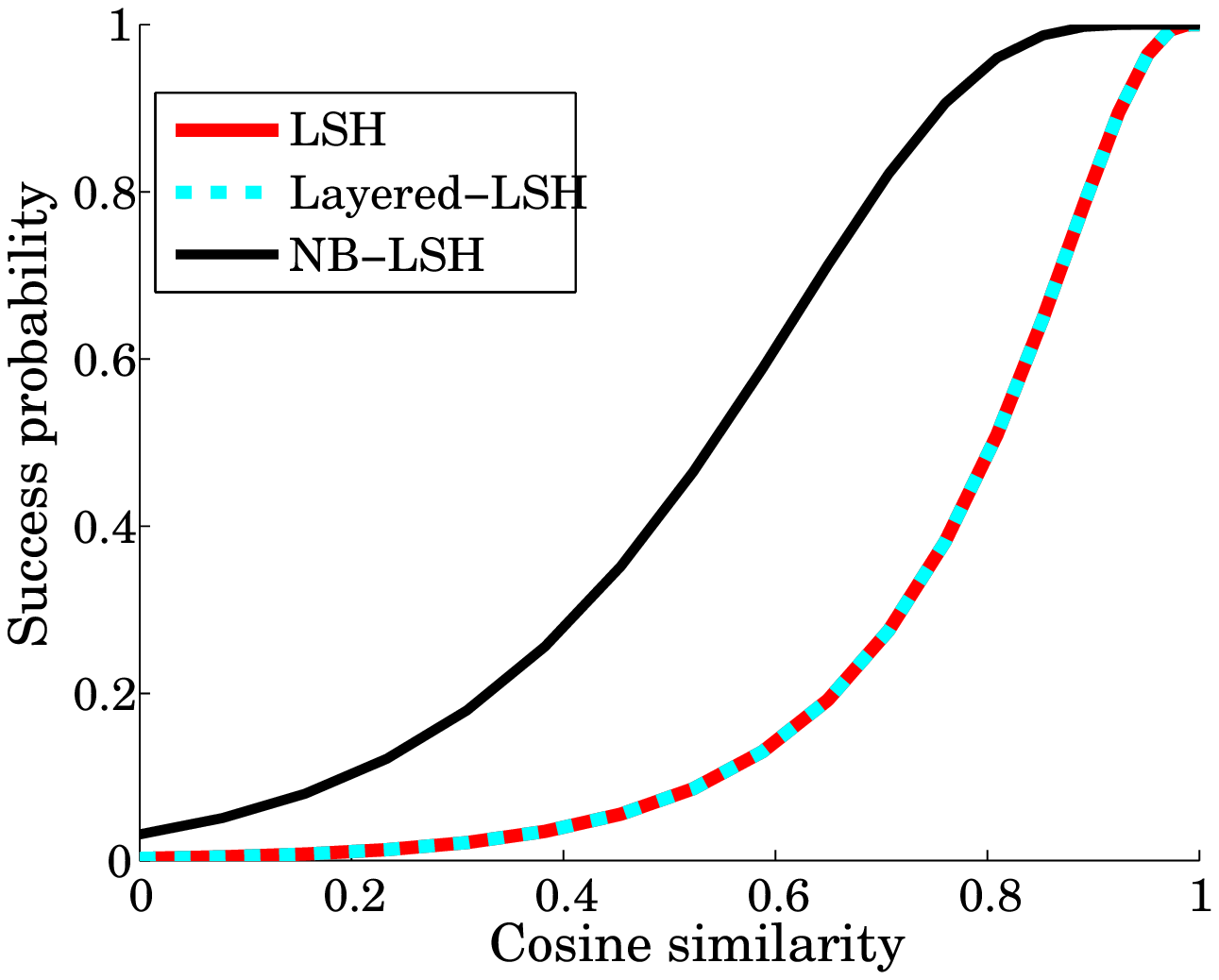}
                \caption{$L=10$}
        \end{subfigure}
        ~ 
        \begin{subfigure}{0.3\textwidth}
                \includegraphics[height=1.3in]{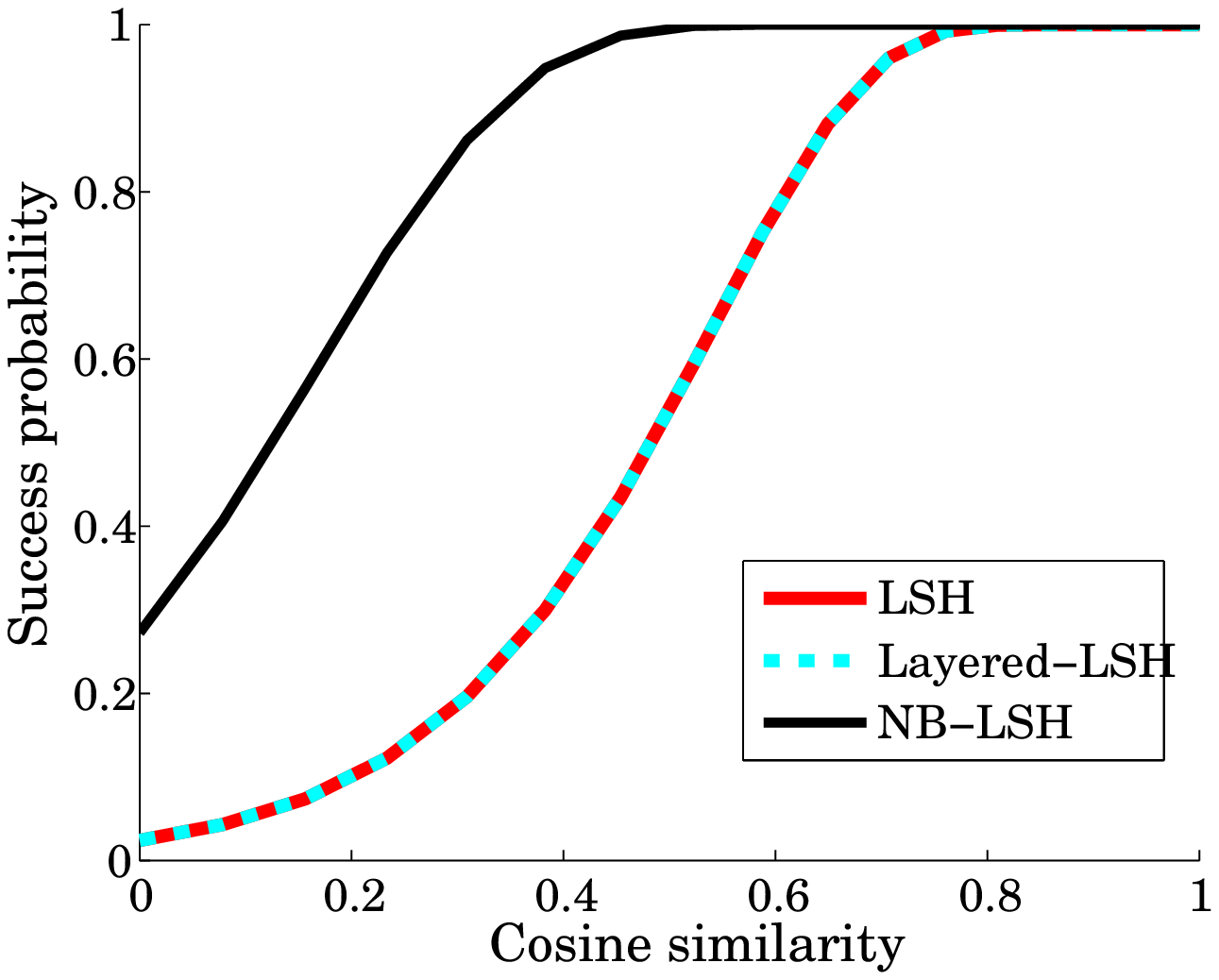}
                \caption{$L=100$}
        \end{subfigure}
        \caption{Analytical success probability as a function of $L$ ($k=12$).
				         NearBucket-LSH guarantees a greater or equal success probability compared to LSH,
								 as it searches in more buckets (namely, near buckets). 
								 The gap increases as $L$ increases.}
								\label{fig:analysis_by_L}
\end{figure*}

\paragraph*{Constant Number of Hash Functions}
\label{sec:nHash}
We compare LSH and NearBucket-LSH for a constant $L$.
Figure \ref{fig:analysis_by_L} depicts their success probability distributions for $k=12$.
As the graphs demonstrate, the success probability of NearBucket-LSH is greater than or equal to the success probability of LSH
for a constant $L$.
This stems from the fact that NearBucket-LSH searches in $kL$ additional near buckets,
which increases its success probability.

\paragraph*{Network Efficiency}
Having seen that NearBucket-LSH increases the success probability for the same $L$
while reducing the effectiveness per bucket,
we proceed to analyze this tradeoff
by comparing \emph{network cost} (average number of messages sent per query) in our concrete CAN-based implementation.
We distinguish between the cached (CNB-LSH) and non-cached (NB-LSH) versions of NearBucket-LSH.
Figure \ref{fig:analysis_searchNetCost} below illustrates that, 
thanks to the low network cost of searching near buckets,
NearBucket-LSH, (and more notably CNB-LSH), 
improves LSH's success probability, for a given average number of messages.

The distributed algorithms contact several bucket nodes, in which they perform local similarity search.
The first column of Table \ref{table:costs} summarizes the number of bucket nodes contacted (and searched) 
by each of the algorithm variants.
NB-LSH contacts $kL$ additional near bucket nodes, which CNB-LSH avoids thanks to caching.

Looking up an exact bucket node requires an average of $k/2$ routing hops in CAN.
Given a query, all algorithms lookup $L$ exact bucket nodes, for an average cost of $\frac{1}{2}kL$
messages per query.
Once reaching an exact bucket node, LSH and CNB-LSH return, 
while NB-LSH forwards messages to its $k$ near bucket nodes.
Contacting a neighbor node in CAN costs one message, for an average of $kL$ additional messages per query.
The second column in Table \ref{table:costs} summarizes the average number of messages per query.

\begin{table*}[hbt]
\begin{center}
\begin{footnotesize}
\begin{tabular}{|c|c|c|c|c|}
\hline
             & Number of nodes     & Average number of  & Number of vectors       & Number of vectors       \\ 
					   & contacted per query & messages per query & stored in a node        & searched per query      \\ \hline \hline
LSH          & L                   & $\frac{1}{2}kL$    & $B$                     & $LB$                    \\ \hline \hline
Layered-LSH  & L                   & $\frac{1}{2}kL$    & $B$                     & $LB$                    \\ \hline \hline
NB-LSH       & L(1+k)              & $1\frac{1}{2}kL$   & $B$                     & $L(k+1)B$               \\ \hline \hline
CNB-LSH      & L                   & $\frac{1}{2}kL$    & $(k+1)B$                & $L(k+1)B$               \\ \hline \hline

\end{tabular}
\end{footnotesize}
\caption{Summary of costs of similarity search in CAN-based LSH variants.}
\label{table:costs}
\end{center}
\end{table*}

Figure \ref{fig:analysis_searchNetCost} shows the success probabilities 
as a function of the average number of messages per query,
for $k=12$ and network costs of $18$, $180$, and $1800$ messages per query.
For each algorithm, we selected the $L$ value that gives us the desired network cost (according to Table \ref{table:costs}).
The graphs demonstrate that NB-LSH increases the success probability of LSH at a low network cost.
CNB-LSH further reduces the network cost, and achieves NB-LSH's success probability, while preserving LSH's network cost.
Note that one could further extend NearBucket-LSH to search in near buckets that differ from the query's bucket in more than one entry.
The success probability of such buckets decreases (Proposition \ref{prop:bNearVS}),
whereas the network cost in NB-LSH and the storage cost in CNB-LSH increases compared to $1$-near buckets.
Thus, searching additional buckets is expected to be less effective.

\begin{figure*}[hbt]
        \centering
        \begin{subfigure}{0.3\textwidth}
                \includegraphics[height=1.3in]{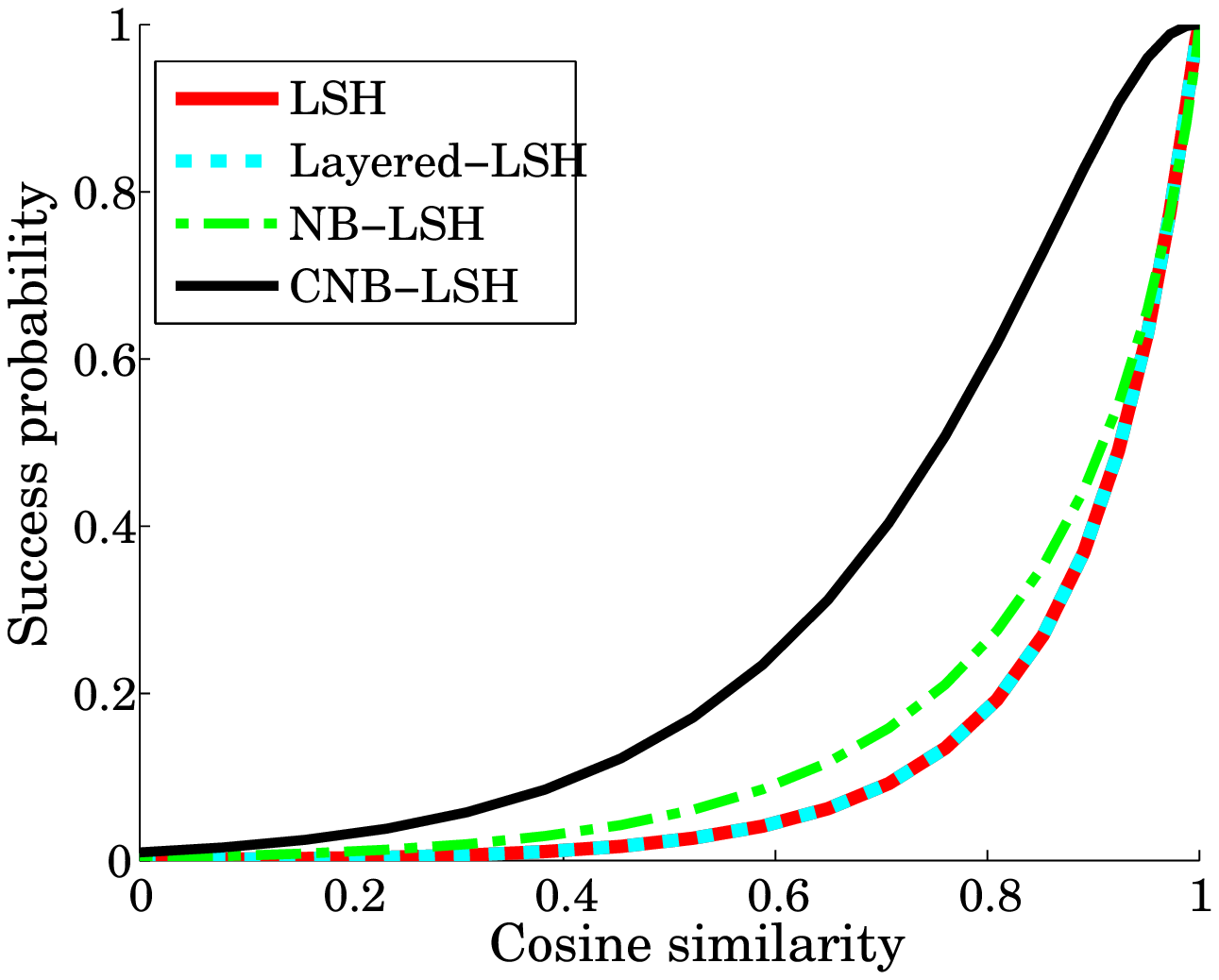}
                \caption{network cost: 18}
        \end{subfigure}%
        ~ 
        \begin{subfigure}{0.3\textwidth}
                \includegraphics[height=1.3in]{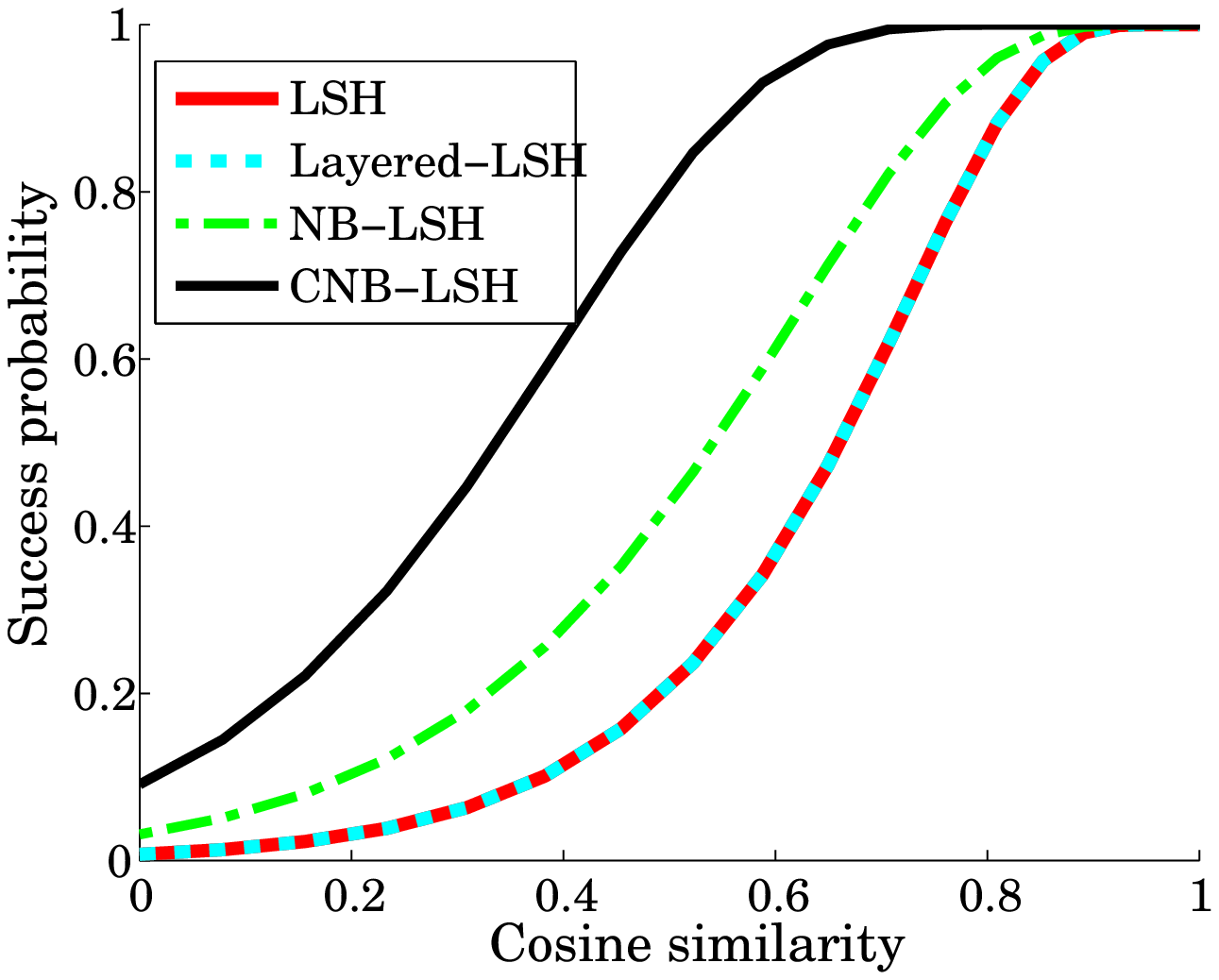}
                \caption{network cost: 180}
        \end{subfigure}
        ~ 
        \begin{subfigure}{0.3\textwidth}
                \includegraphics[height=1.3in]{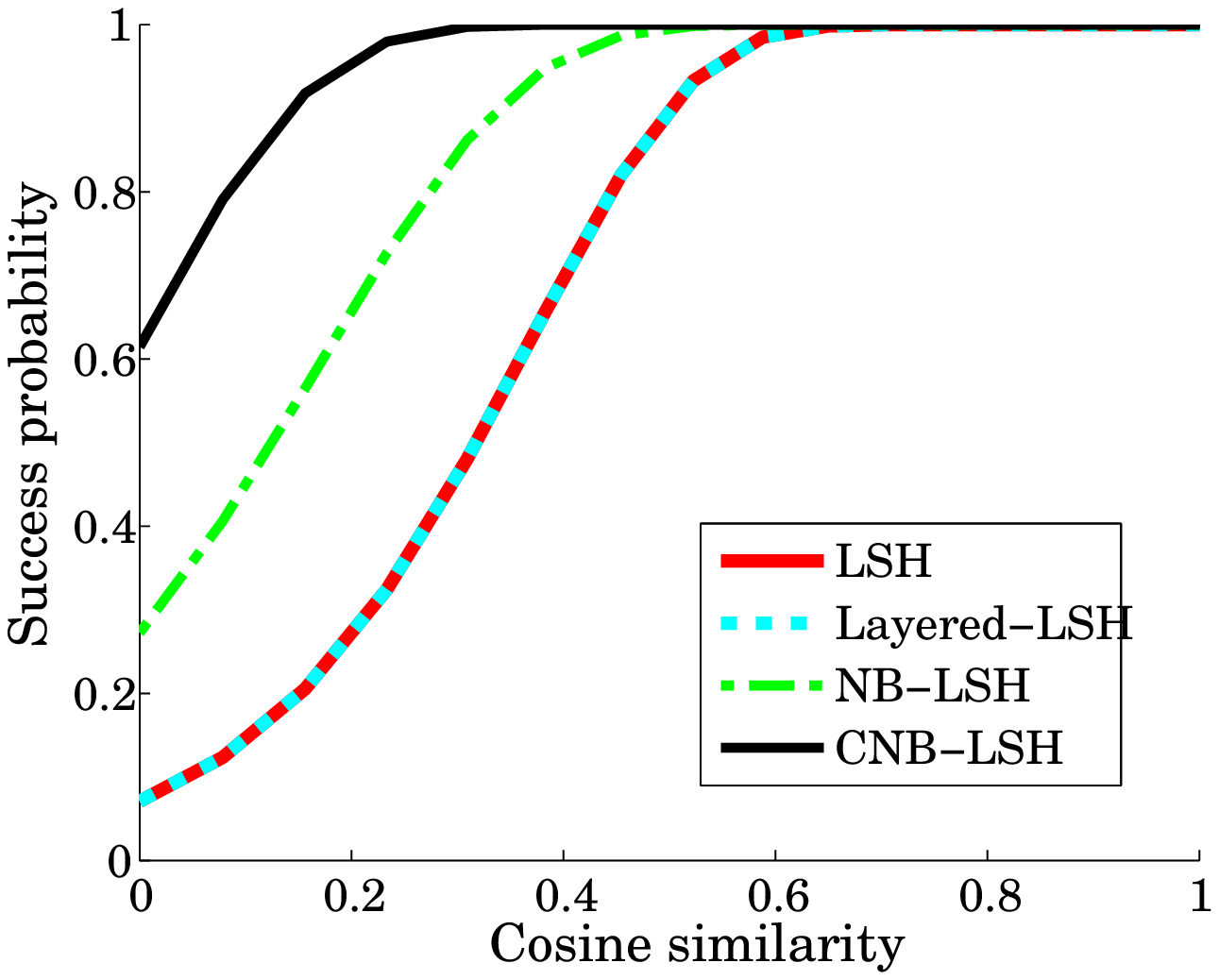}
                \caption{network cost: 1800}
        \end{subfigure}
        \caption{Analytical success probability as a function of network cost ($k=12$).
				         NB-LSH exploits the low lookup cost of near buckets in CAN, 
								 and increases LSH's success probability for a low network cost.
								 CNB-LSH further saves messages by caching near buckets,
								 and achieves the greatest success probability for a given network cost.}
								 \label{fig:analysis_searchNetCost}
\end{figure*}

\paragraph*{Other considerations}
Our work focuses on minimizing the network cost, which is a dominant cost in P2P networks.
For completeness, we present in the third and fourth columns of Table \ref{table:costs} 
other costs which tradeoff with network-efficiency.
We denote the average bucket size by $B$.
In terms of storage capacity, NB-LSH preserves the same space complexity as LSH.
CNB-LSH increases the space complexity due to caching,
while being more network-efficient than NB-LSH.
Both NearBucket-LSH variants search over a larger number of vectors than LSH,
implying more processing work per query.
As our algorithm searches the buckets in parallel, 
and the average bucket size is equal in all algorithms,
this does not affect the query latency.

\section{Evaluation}
\label{sec:eval}
We previously showed that for a fixed network cost, 
CNB-LSH has the greatest probability  
to find an object $y$ that is $s$-similar to a given query,
for any similarity value $s$.
Given this theoretical property, 
we expect CNB-LSH to demonstrate the best search quality for real data
and commonly used search quality metrics.

In this section, we empirically evaluate the algorithm.
In Section \ref{sec:measures}, we formally define the search quality metrics that we use,
and in Section \ref{sec:methodology}, we detail the evaluation methodology.
We move on to elaborate on our experiments and results.
In Section \ref{sec:val}, we empirically reproduce the theoretical success probability analysis of Section \ref{sec:analysis}.
In Section \ref{sec:qResults}, we valuate LSH, Layered-LSH, NB-LSH, and CNB-LSH on three real world OSN datasets of varying sizes.

\subsection{Measures}
\label{sec:measures}
We measure the network cost by the average number of messages per query,
according to Table \ref{table:costs}.
As we deal with approximate similarity search,
we measure an algorithm's search quality by recall and precision.

\paragraph*{Recall}
\label{sec:recall}
(at $m$) is defined as follows~\cite{Lv07}:
\begin{defn}[recall at $m$]
Given a query $q$, let $I_m(q)$ denote its ideal $m$-result set.
Let $A_m(q)$ denote the approximate $m$-result set of $q$ returned by some algorithm $A$.
The recall is the fraction of results from the $m$-ideal result set that are returned by $A$:
\begin{equation}
recall@m(A,q) = \frac{|A_m(q) \cap I_m(q)|}{|I_m(q)|}.
\end{equation}
\end{defn}
A high recall indicates that algorithm $A$ approximates well the ideal $m$-result set of query $q$.
A value of $1$ is optimal.
An algorithm's recall is the mean of the queries' recall averaged over a query set $Q$:
\begin{defn}
\begin{equation}
recall@m\left(A\right) = \frac{1}{|Q|}\sum_{q \in Q} recall@m(A,q).
\end{equation}
\end{defn}

\paragraph*{Normalized Cumulative Similarity}
\label{sec:ncs}
In order to measure the precision of an approximate similarity search algorithm for a query $q$,
we compare the similarities of its $m$-result set to those of the ideal $m$-result set.
We do this by defining the following ratio, which we name the \emph{normalized cumulative similarity (NCS)}:
\begin{defn}[NCS at $m$]
Given a query $q$,\\ let $CumSim(I_m,q)$ denote the sum of the similarity values to $q$ of the results in the ideal $m$-result set.\\
Let $CumSim(A_m,q)$ denote the sum of the similarity values to $q$ of the results in the $m$-result set of a given algorithm $A$.
Then,
\begin{equation}
NCS@m(A,q) = \frac{CumSim(A_m,q)}{CumSim(I_m,q)}
\end{equation}
\end{defn}
Note that $CumSim(I_m,q) \ge CumSim(A_m,q)$, and both are positive.
Therefore, $NCS@m(A,q) \in [0,1]$.

One may think of NCS as the ratio between the average similarity of the retrieved (approximate) result set,
and the average similarity of the ideal result set.
The closer NCS is to $1$, the more precise the approximate result set is.

We measure the NCS of an algorithm by averaging it over the query set $Q$:
\begin{equation}
NCS@m(A) = \frac{1}{|Q|}\sum_{q \in Q} NCS@m(A,q)~.
\end{equation}
 
\subsection{Methodology}
\label{sec:methodology}
\paragraph*{Datasets}
We use three real-world publically-available datasets of OSNs~\cite{Yang2012}:

\begin{itemize}
	\item \emph{DBLP}~\cite{DBLP}, the computer science bibliography database:
Authors are users, and venues are interests.
We use a crawl of $13,\!477$ interests, and $260,\!998$ users that have at least one interest.

\item \emph{LiveJournal}~\cite{Livejournal} blogging-based OSN:
Users publish blogs and form interest groups, which users can join.
The LiveJournal crawl consists of $664,\!414$ such groups, which we consider as user interests. 
There are $1,\!147,\!948$ users with at least one interest.

\item \emph{Friendster}~\cite{Friendster} online gaming network:
Similarly to LiveJournal, Friendster allows users to form interest groups, which we consider as interests.
The dataset consists of $1,\!620,\!991$ interest groups, and $7,\!944,\!949$ users with at least one interest. 
\end{itemize}

All datasets contain anonymous user ids and interest information.
We filtered out users having no interest.

\paragraph*{Parameters}
We set $k=10$ in DBLP, $k=12$ in LiveJournal and $k=15$ in Friendster.
We follow previous art~\cite{Bahmani12,Haghan09} that uses $k$ values between $10$ and $20$,
and bucket sizes of a few hundreds~\cite{Gionis99}.
Thus, we have $1,\!024$ buckets in DBLP, $4,\!096$ in LiveJournal, and $32,\!768$ in Friendster.
The average bucket size is approximately $250$ vectors in all datasets.
We set $m$, the number of search results, to $10$.

\paragraph*{Creating Sketch Vectors}
\label{sec:DistributedIndexingSimulation}
We construct users' weighted interest vectors according to the dataset at hand.
We weight each interest $I$ based on its inverse frequency in user vectors~\cite{Adamic01friendsand}:
$w(I) = ln(\frac{N_u}{N_I+1})+1$, where $N_u$ denotes the total number of users,
and $N_I$ denotes the number of users having interest $I$.
The user vector entry $v_i$ is zero or $w(I)$ according to whether the user is associated with specific interest $I$.
We use TarsosLSH's~\cite{TarsosLSH}
cosine-LSH implementation for mapping user vectors into LSH buckets (given $k$ and $L$ parameters),
which we modify to be more memory-efficient (based on Java HashMap),
in order to support our large and sparse user vectors.
For mapping buckets into peers in Layered-LSH, 
we implement a Hamming-based LSH according to \cite{Gionis99,Chierichetti12},
which is absent in the TarsosLSH library we use.

\paragraph*{Simulator}
\label{sec:simulator}
We implement a simulator of our CAN-based overlay
using Apache Lucene 4.3.0~\cite{Lucene} centralized search index.
We simulate distributing user vectors in bucket nodes
by indexing vectors by their hash values (sketch vectors).
The hash is then used for looking up a specific bucket node, 
and local similarity search is performed by limiting the search to the selected bucket
(using Lucene's Filter mechanism).
We compute the cosine similarity according to Equation \ref{eq:cos}\footnote{Lucene's similarity function is very close to cosine similarity though not identical. Therefore, in order to retrieve the cosine-based top-10 results, we retrieve top-100 results using Lucene's scoring mechanism, and then re-rank them using cosine.}
for Lucene's top $100$ results in each of the searched bucket,
and return the top $m$.
If multiple buckets are searched, the results are then merged and the top $m$ are returned.
To compute the ideal result set of a given query, 
we use Lucene to extract the top $100$ results for the entire dataset, and return the top $m$.

\subsection{Success Probability}
\label{sec:val}
Our first experiment reproduces the analytical results of Section \ref{sec:analysis}.
The success probability equations therein express the probability to find a random vector $y$
in any of the searched buckets.
As our algorithms only return the top $m=10$ results,
there is a low probability that they return a random $y$,
even if it is in one of the searched buckets.
We therefore only evaluate the success probability for ideal top results, which,
if present in the searched bucket, are returned by our algorithms.

We randomly sample a set of queries and construct a corresponding set of pairs $(x,y)$,
where $x$ is a random query vector, and $y$ is its top result in the ideal set.
We define the similarity intervals: $[0,0.1), [0.1, 0.2), \cdots, [0.9,1]$.
We associate with each interval the pairs $(x,y)$ for which
the similarity between $x$ and $y$ falls within that interval. 
For each interval, we measure the fraction of its pairs for which an algorithm successfully finds $y$.
We conduct the experiment using $3,\!000$ pairs.
In the DBLP sample, there are no $y$ values with low scores,
and therefore these intervals are missing from the empirical graph.

\begin{figure*}[hbt]
        \centering	
		
	\begin{subfigure}{0.3\textwidth}
                \includegraphics[height=1.3in]{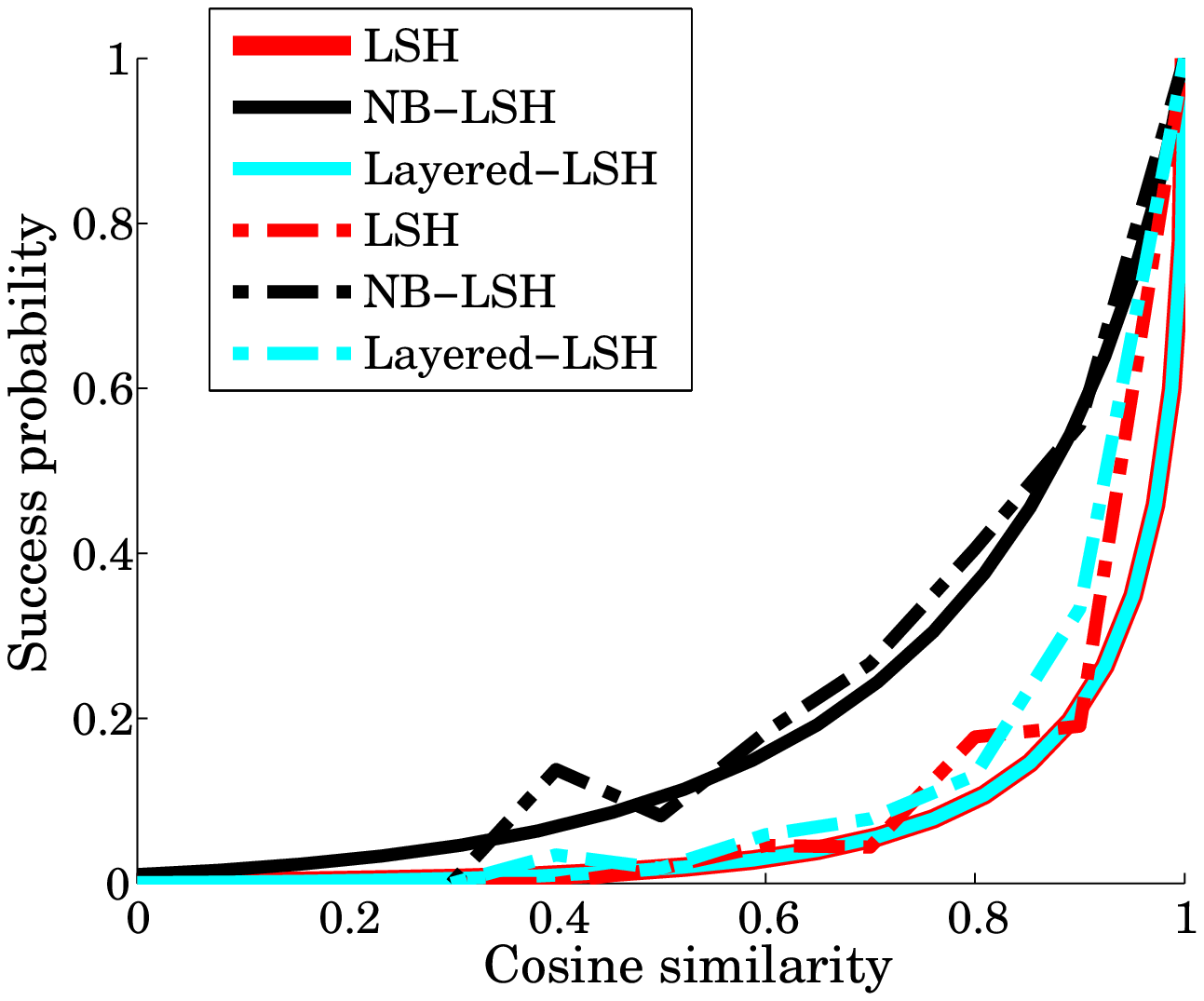}
                \caption{DBLP $L=1$}
    \end{subfigure}%
					\begin{subfigure}{0.3\textwidth}
						\includegraphics[height=1.3in]{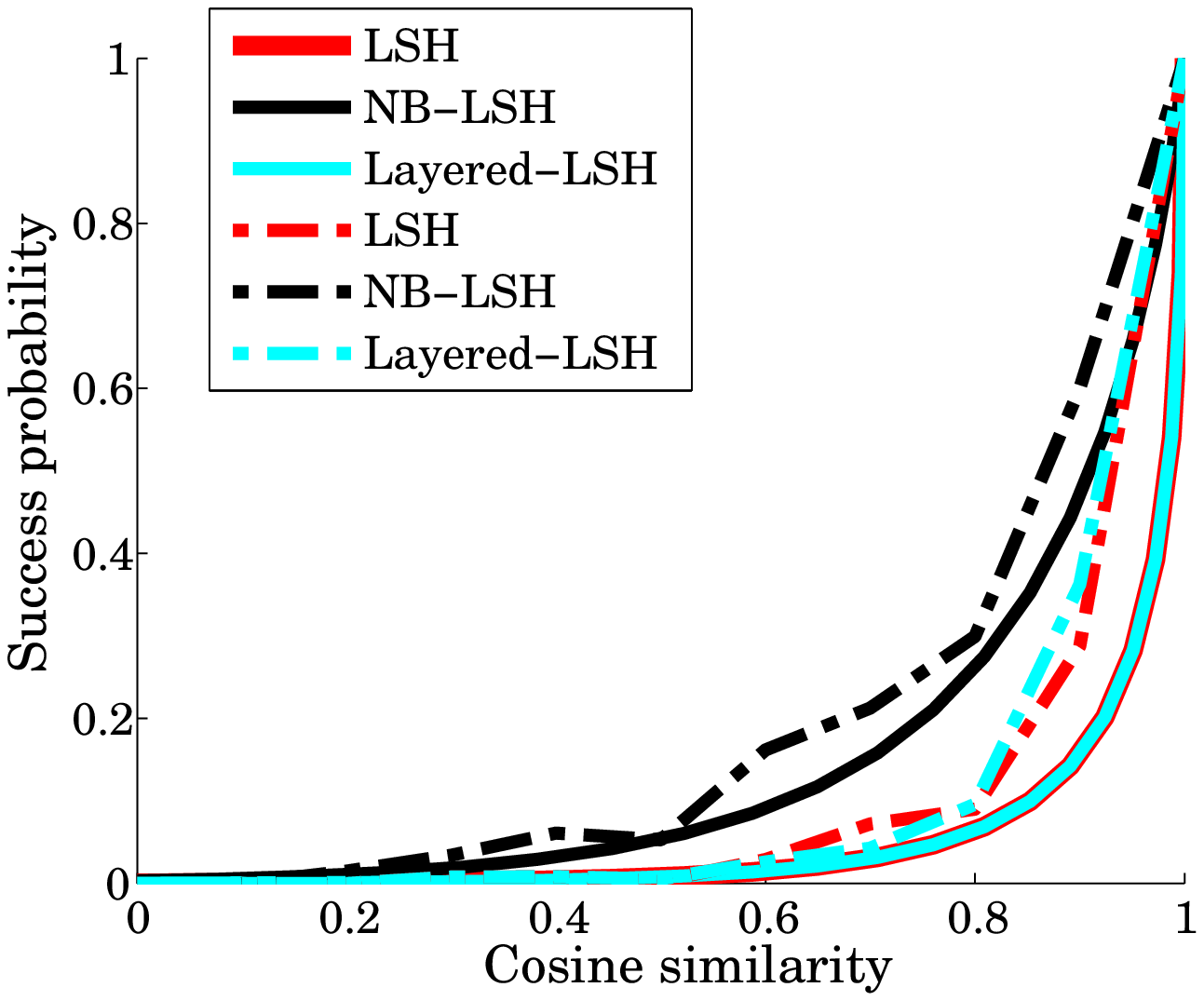}
            \caption{LiveJournal $L=1$}
        \end{subfigure}%
        ~ 
						\begin{subfigure}{0.3\textwidth}
						\includegraphics[height=1.3in]{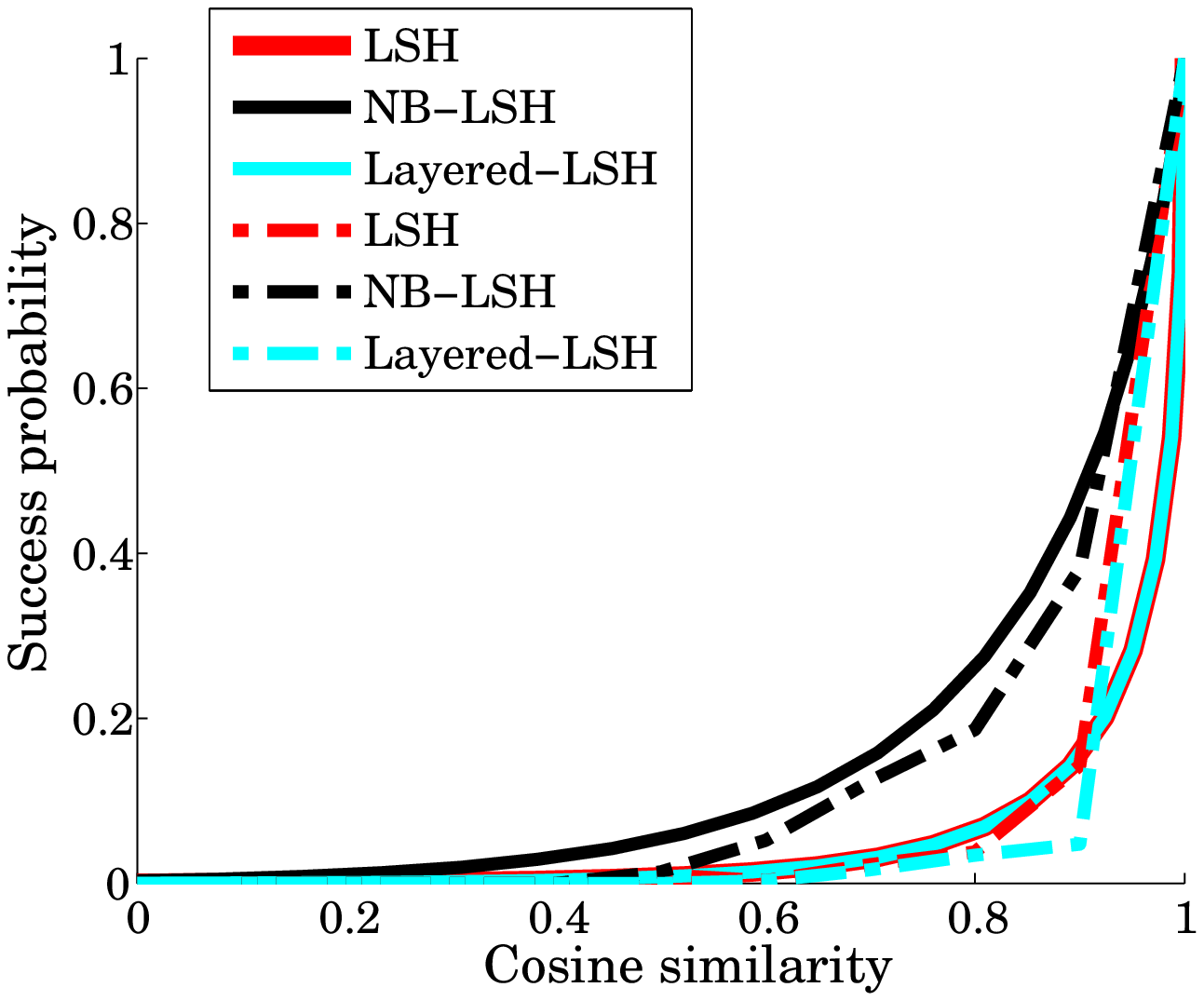}
            \caption{Friendster $L=1$}
        \end{subfigure}%
        ~ 
		
        \begin{subfigure}{0.3\textwidth}
                \includegraphics[height=1.3in]{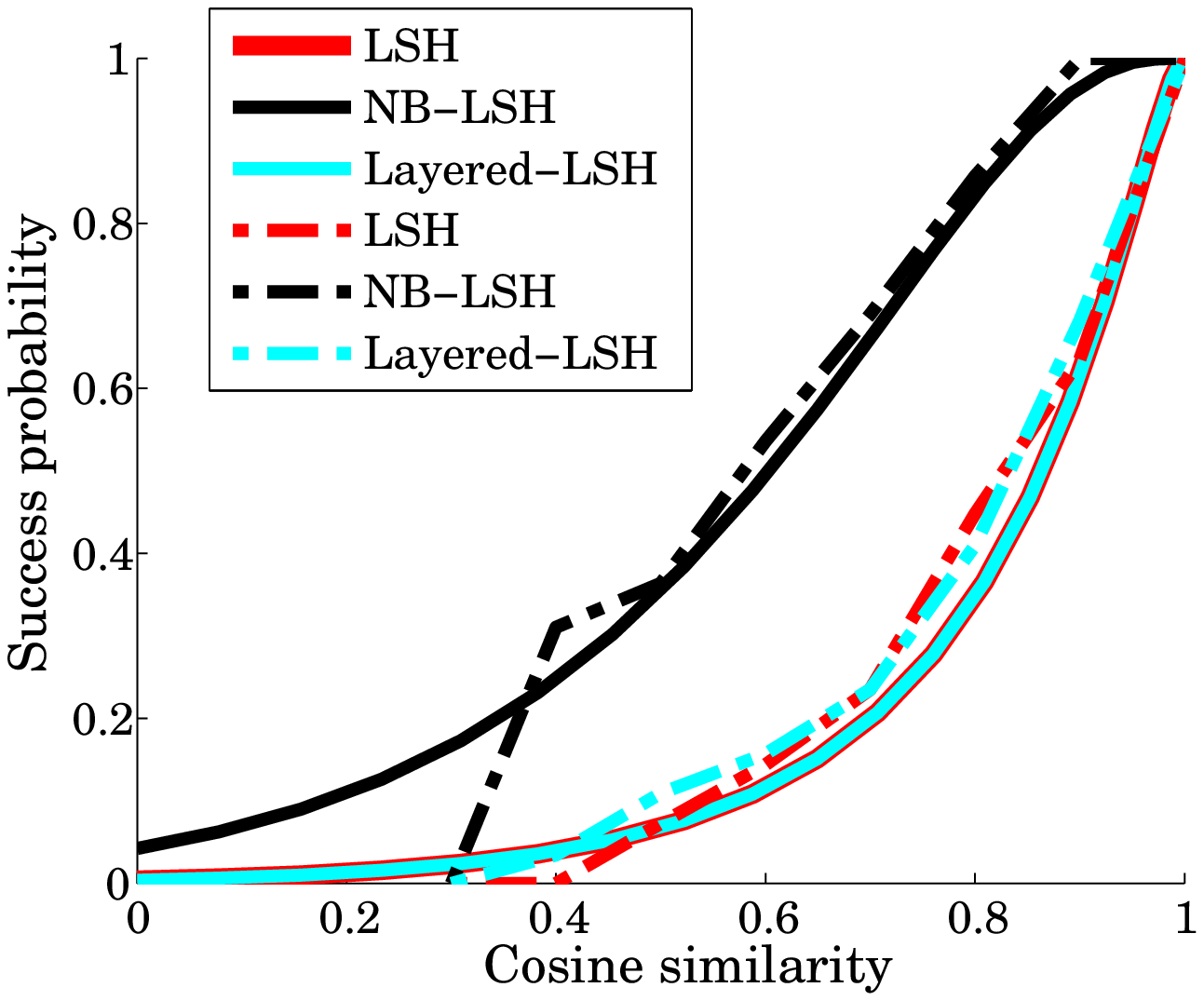}
                \caption{DBLP $L=4$}
        \end{subfigure}
        ~ 
        \begin{subfigure}{0.3\textwidth}
                \includegraphics[height=1.3in]{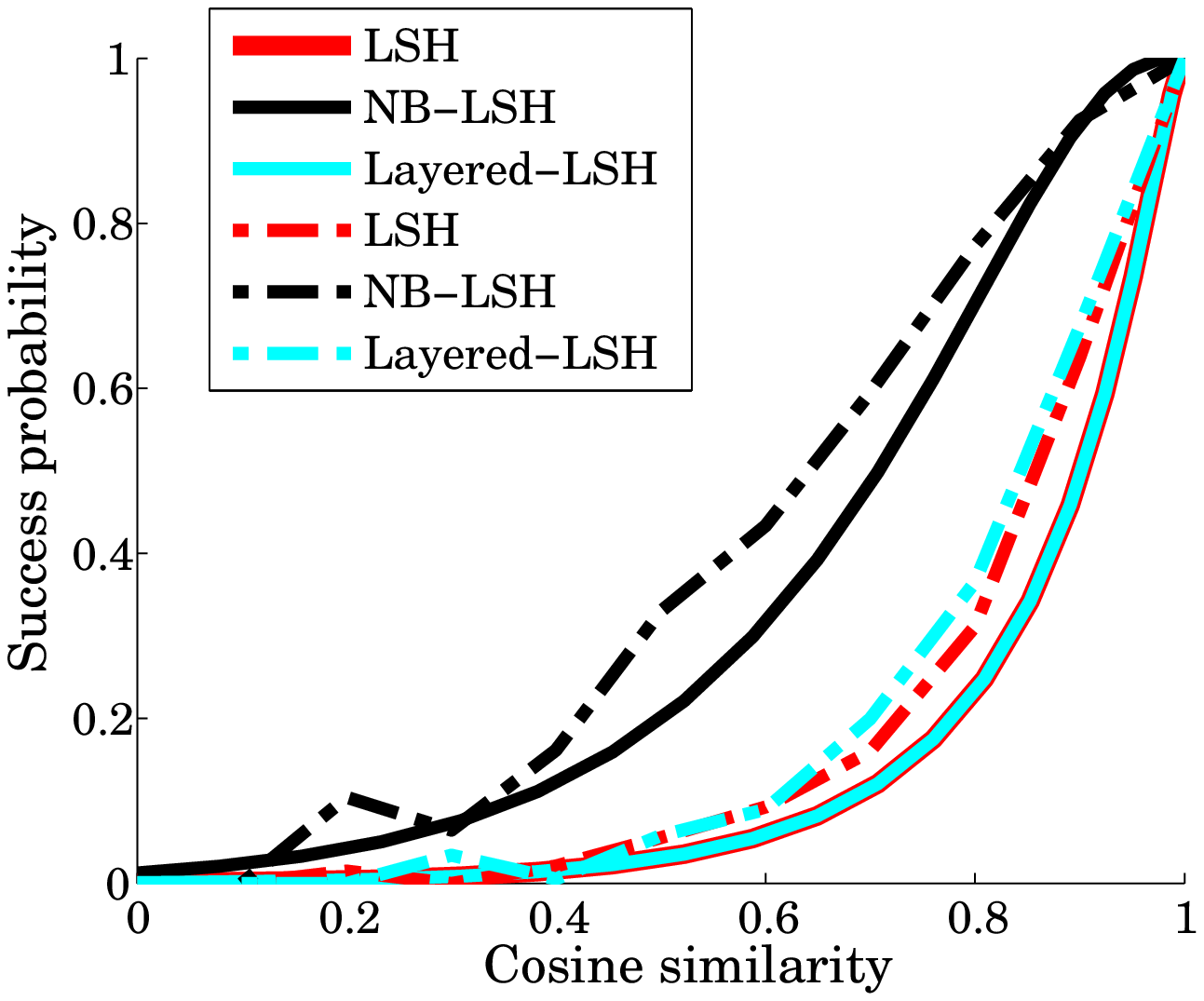}
                \caption{LiveJournal $L=4$}
        \end{subfigure}
        ~ 
        \begin{subfigure}{0.3\textwidth}
                \includegraphics[height=1.3in]{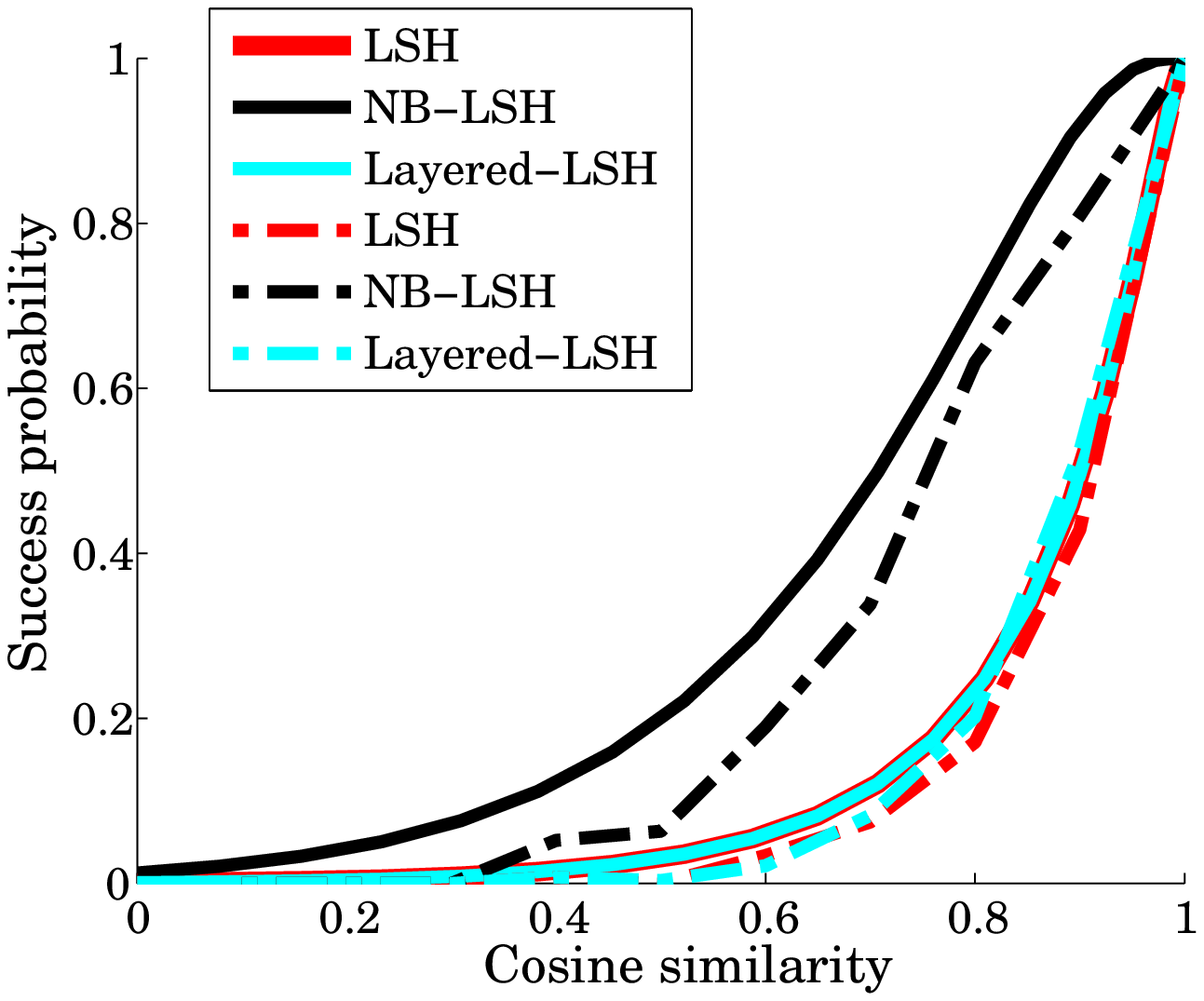}
                \caption{Friendster $L=4$}
        \end{subfigure}

        \caption{Analytical (solid) vs. observed (dotted) success probability in DBLP,
				        LiveJournal ($k=12$), and Friendster ($k=15$).
								The observed success probability follows the trend of the analytical success probability.}
								\label{fig:analysis_vs_empirical}
\end{figure*}

Figure \ref{fig:analysis_vs_empirical} depicts the analytical success probabilities 
(solid, computed according to Propositions \ref{prop:lsh} and \ref{prop:nblsh}),
and the success probabilities we observed in our experiment (dotted).
The empirical graphs follow the trend of the analytical graphs with some estimation error.
The estimation error stems from two reasons:
(1) sampling error (the average number of samples in an interval is $300$), and
(2) the sampling of $y$ is biased as we only select the most similar vector to $x$.

\subsection{Search Quality Results}
\label{sec:qResults}
Having looked at the success probability of the top result,
we now turn to study search quality for the top $m=10$ results. 
In order to measure search quality,
we construct a query set of $3,\!000$ randomly sampled users.
For each query $q$, we compute its ideal result set,
as well as the result sets according to the algorithms we compare.
For each dataset, we measure recall and precision over the query set in use.

\begin{figure*}[hbt]
        \centering
				\begin{subfigure}{0.3\textwidth}
								\includegraphics[width=\textwidth]{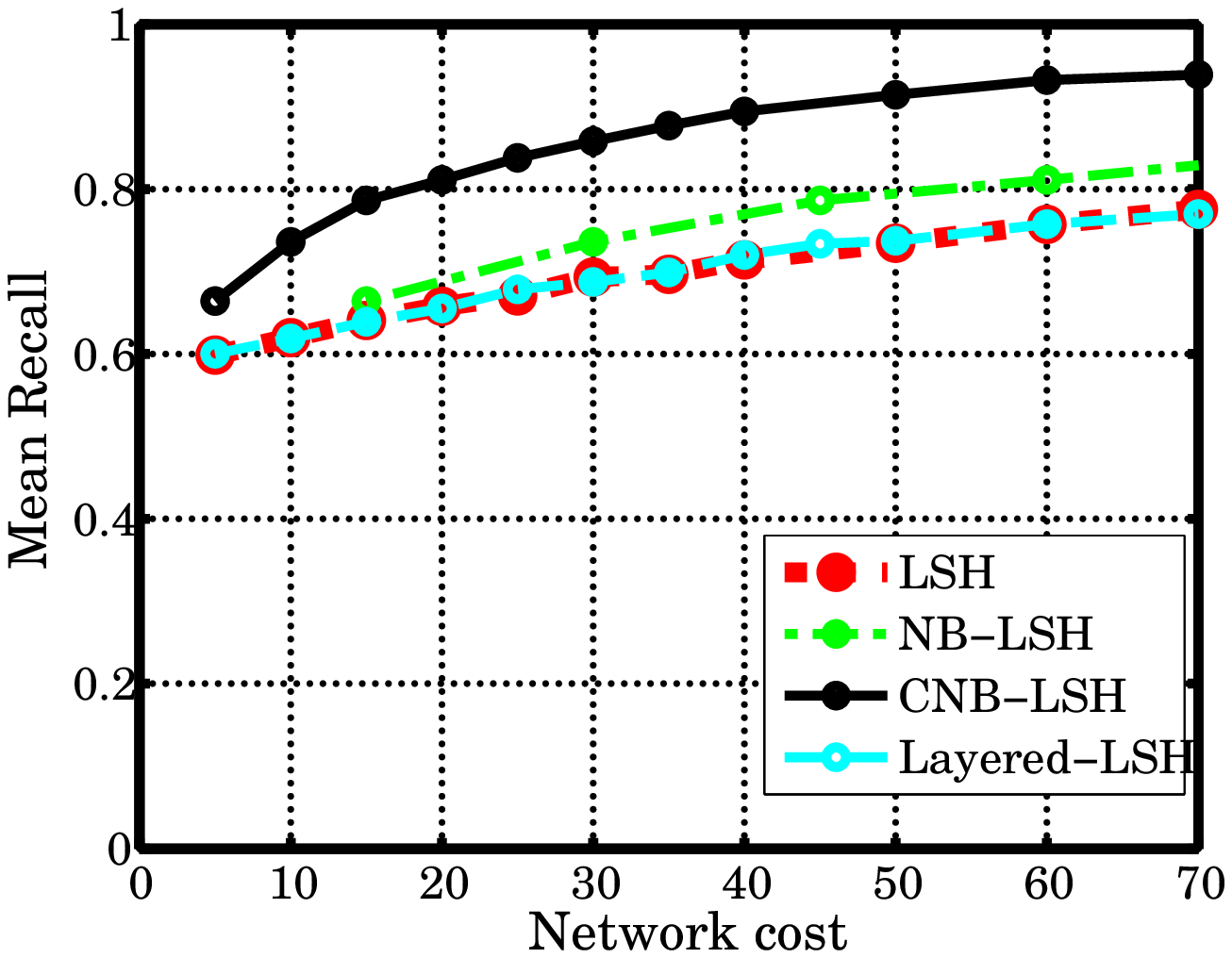}							
                \caption{DBLP recall}
        \end{subfigure}
				~ 
				\begin{subfigure}{0.3\textwidth}
					\includegraphics[width=\textwidth]{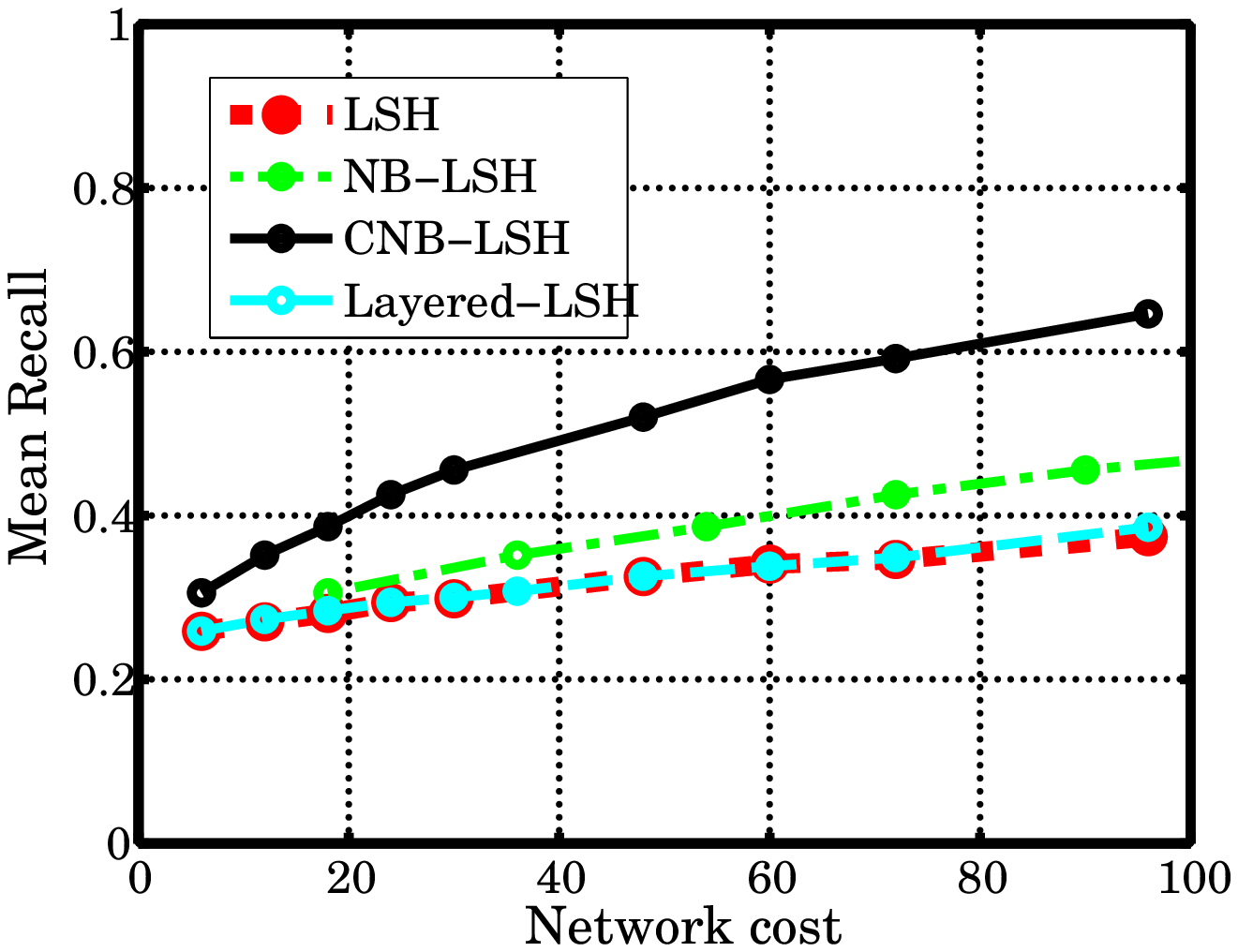}
					\caption{LiveJournal recall}
        \end{subfigure}
				~ 
				\begin{subfigure}{0.3\textwidth}
					\includegraphics[width=\textwidth]{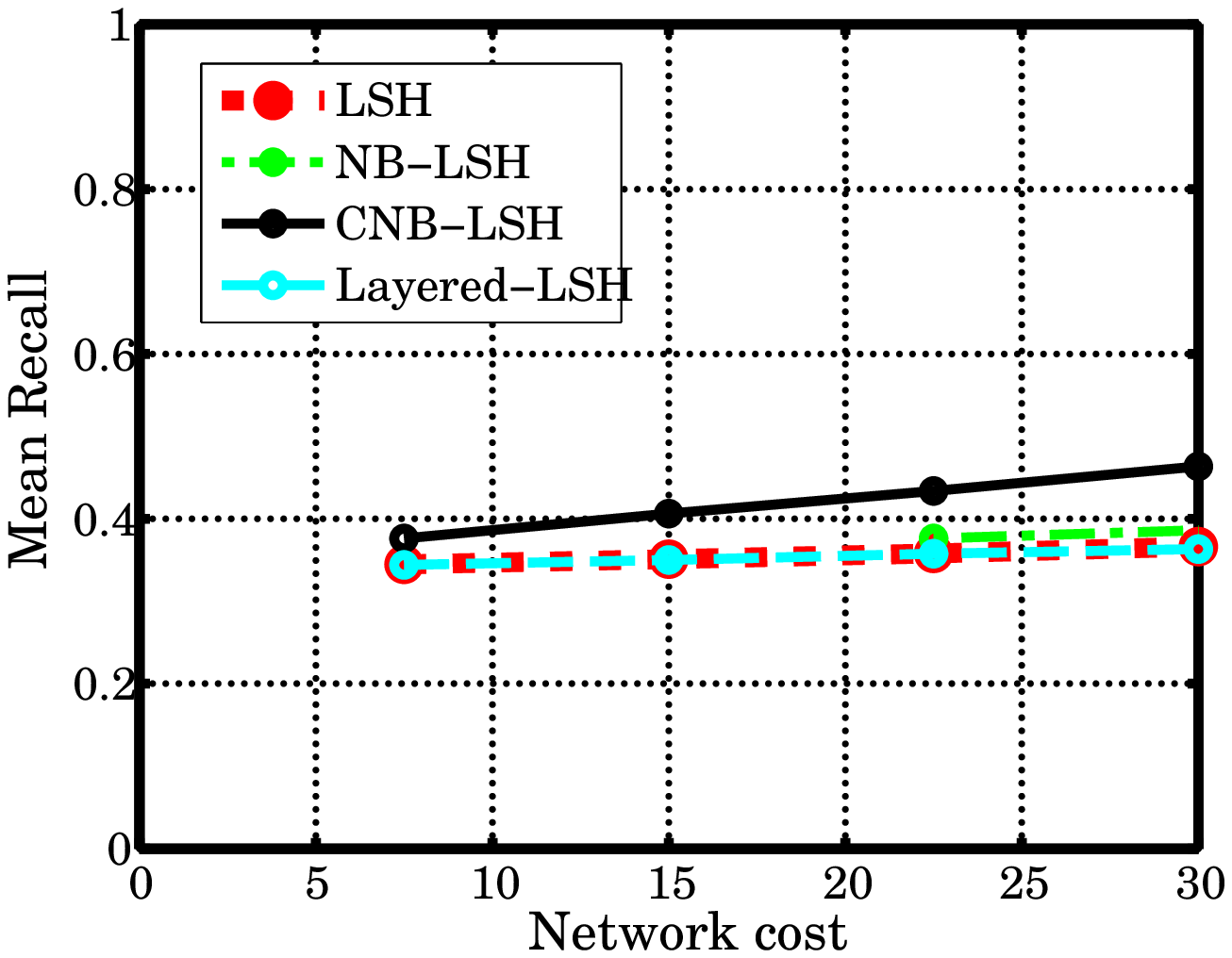}
           \caption{Friendster recall}
        \end{subfigure}

        \begin{subfigure}{0.3\textwidth}
                \includegraphics[width=\textwidth]{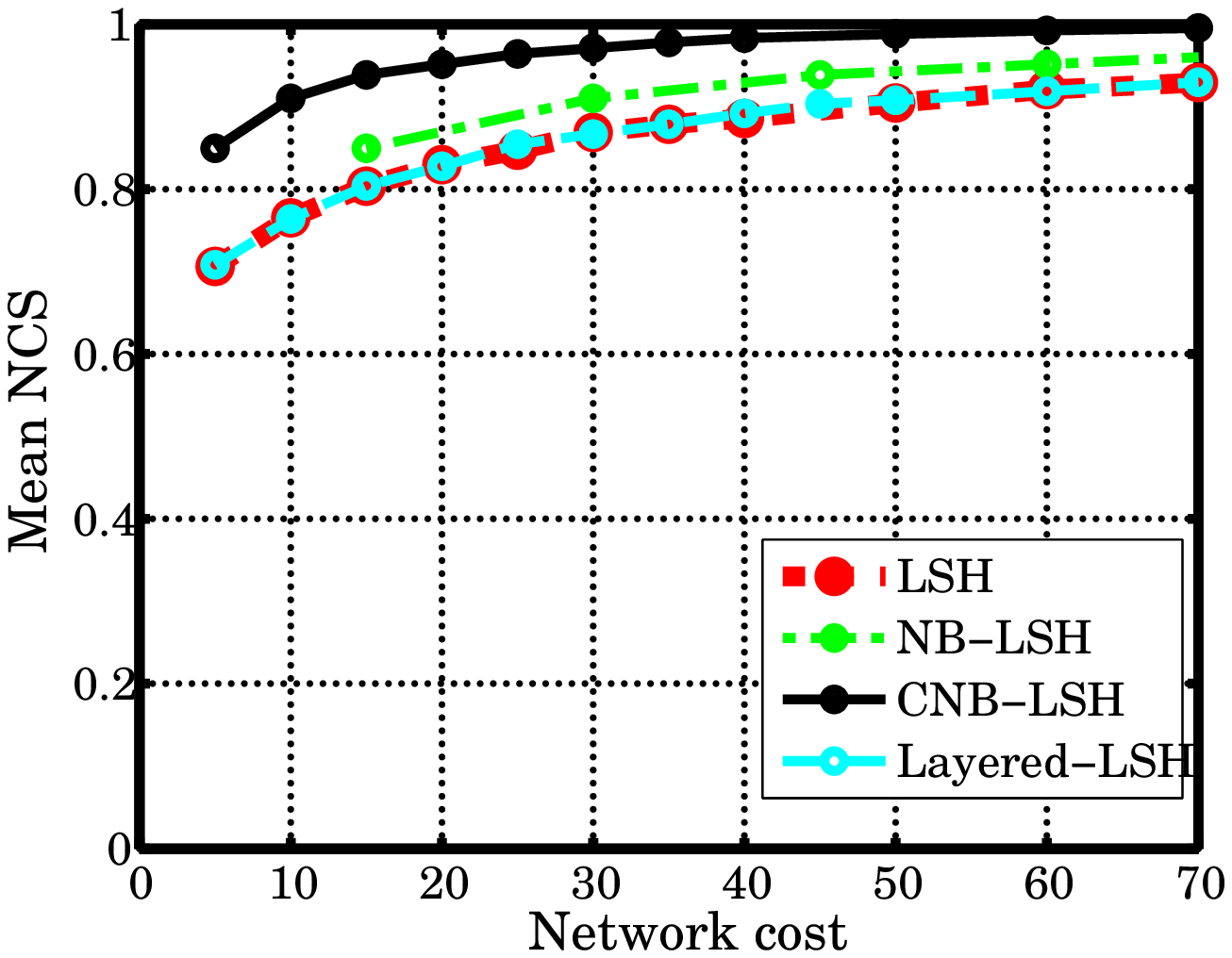}
                \caption{DBLP precision}
        \end{subfigure}
				~ 
        \begin{subfigure}{0.3\textwidth}
                \includegraphics[width=\textwidth]{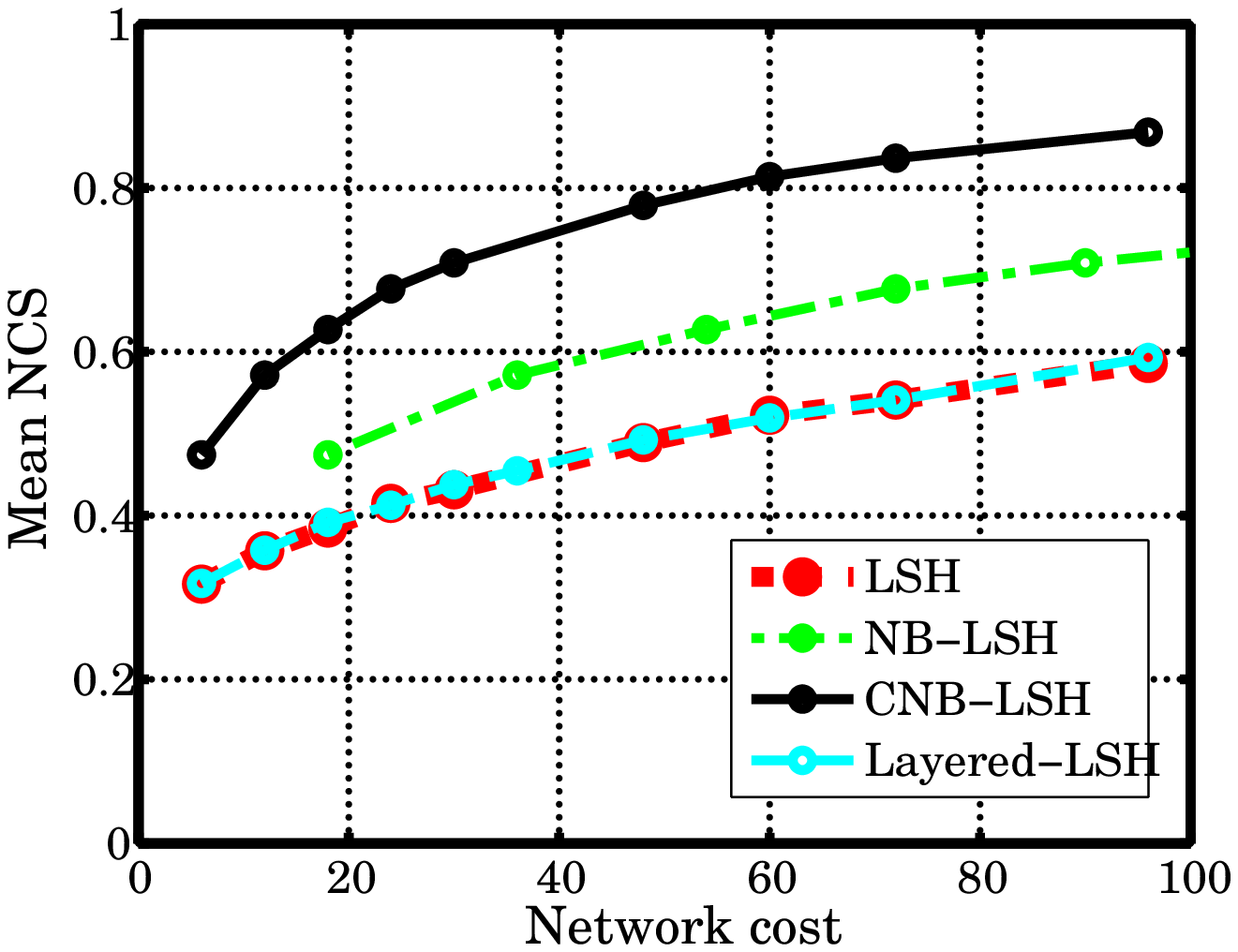}
                \caption{LiveJournal precision}
        \end{subfigure}
				~ 
        \begin{subfigure}{0.3\textwidth}
                \includegraphics[width=\textwidth]{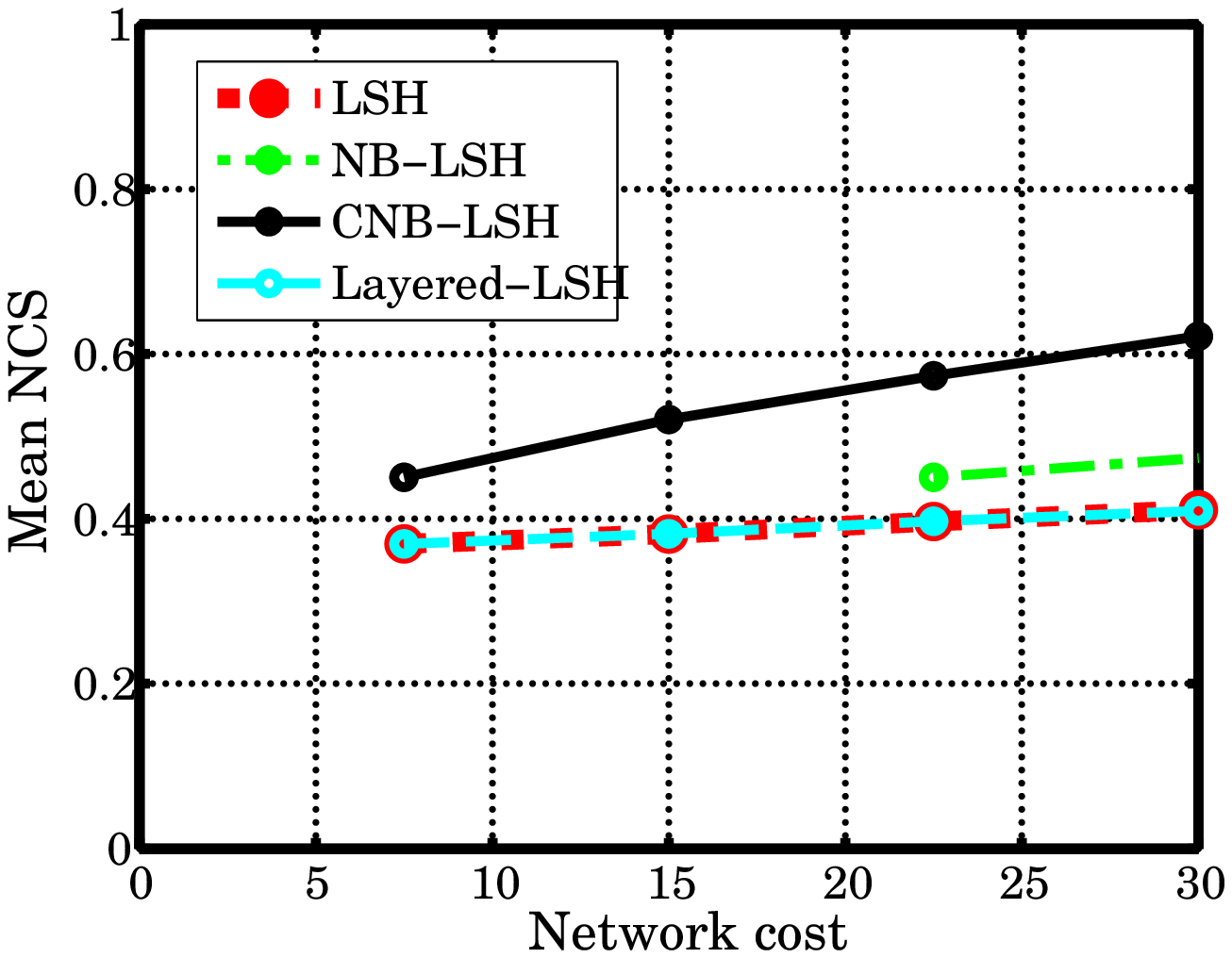}
                \caption{Friendster precision}
        \end{subfigure}
				~ 
         \caption{Search quality as a function of the average number of messages per query,
				         for three real world datasets: DBLP, LiveJournal, and Friendster
				        ($k=10$, $k=12$, $k=15$, respectively).
								 For all datasets, CNB-LSH provides the greatest search quality as a function of the network cost,
								 according to two metrics: recall and precision.}
				\label{fig:searchNetCost_searchQuality}	
\end{figure*}

Figure \ref{fig:searchNetCost_searchQuality} illustrates our experimental results as a function of network cost.
We increase the network cost by gradually increasing $L$,
which increases search quality for all datasets as expected.
We use larger values of $k$ for larger datasets in order to preserve a common average bucket size.
This ensures that local search takes the same time, and the cache sizes are identical.
According to Equations \ref{prop:exBucket} and \ref{prop:nbBucket},
the larger $k$ is, the lower the success probability is, regardless of whether we search in exact or near buckets.
Thus, we expect a decrease in search quality when the dataset size increases,
which is indeed demonstrated in the graphs.

The three datasets show a similar trend.
Layered-LSH's search quality equals that of the basic LSH as expected.
NearBucket-LSH (both cached and non-cached) demonstrates an increase in search quality compared to LSH and Layered-LSH,
which is achieved by searching in additional near buckets stored at neighboring nodes or the node itself.
For example, in LiveJournal (second column), 
LSH requires an average of $96$ messages per query in order to achieve $0.59$ precision,
whereas CNB-LSH achieves a precision of $0.57$ using only $12$ messages.
CNB-LSH also improves recall significantly,
for example, achieving a $0.59$ recall using $72$ queries,
compared to a recall of $0.35$ for LSH.
In all cases, NB-LSH is between LSH and CNB-LSH.

\section{Conclusions and Future Work}
\label{sec:conc}
We presented NearBucket-LSH, a network-efficient LSH algorithm for P2P OSNs,
which provides good search quality.
We first analytically showed that, for cosine similarity, 
our choice of searched near buckets is optimal, that is,
near buckets that differ in a single entry from the query's bucket 
are more likely to contain similar vectors than other near buckets. 
We then showed, both mathematically and empirically,
that one may dramatically lower the additional network cost for searching in these buckets
by exploiting CAN's internal structure and judicious caching.

Our work raises some questions that would be of interest to study in future work.
In several recommendation applications vectors having negative weights are considered.
It may be of interest to explore near buckets search in this context.
In the context of OSNs, it may be interesting to consider other signals,
such as social relations, for improving similarity search.

\bibliographystyle{abbrv}
\bibliography{NearBucket}  



\end{document}